\documentclass[11pt,a4paper]{article}

\usepackage{latexsym}
\usepackage{graphics}
\usepackage{amsmath,amsfonts,amssymb}
\usepackage[utf8]{inputenc}
\usepackage[T1]{fontenc}
\usepackage{esint}
\usepackage{epsfig}
\usepackage{float}
\usepackage{hyperref}
\usepackage{color,xcolor}

\def\NO{\nonumber}

\newcommand{\be}{\begin{equation}}
\newcommand{\ee}{\end{equation}}

\def\bea{\begin{eqnarray}}
\def\eea{\end{eqnarray}}

\def\beqx{\begin{displaymath}}
\def\eeqx{\end{displaymath}}

\newcommand{\bmat}{\left(\begin{array}}
\newcommand{\emat}{\end{array}\right)}

\newtheorem{lemma}{Lemma}[section]

\def\a{\alpha}
\def\b{\beta}

\def\d{\delta}
\def\e{\epsilon}
\def\f{\phi}
\def\g{\gamma}
\def\h{\eta}
\def\i{\iota}
\def\j{\psi}
\def\k{\kappa}

\def\m{\mu}

    \def\om{\omega}
\def\p{\pi}

    \def\th{\theta}

\def\x{\xi}

\def\D{\Delta}

\def\G{\Gamma}

\def\L{\Lambda}

    \def\Om{\Omega}
\def\P{\Pi}

\def\vf{\varphi}

\def\cn{{\cal N}}
\def\co{{\cal O}}
\def\cp{{\cal P}}

\def\cs{{\cal S}}

\def\bo{{\raise-.3ex\hbox{\large$\Box$}}}               
\def\pa{\partial}                                       
\def\face{{\raise.2ex\hbox{$\displaystyle \bigodot$}\mskip-2.2mu \llap {$\ddot
        \smile$}}}                                   
\def\>{\rangle}                                      
\def\<{\langle}                                      

\newcommand{\sub}[1]{\phantom{}_{(#1)}\phantom{}}    
\def\wt#1{\widetilde{#1}}                            
\def\Hat#1{\widehat{#1}}                             
\def\leftrightarrowfill{$\mathsurround=0pt \mathord\leftarrow \mkern-6mu
        \cleaders\hbox{$\mkern-2mu \mathord- \mkern-2mu$}\hfill
        \mkern-6mu \mathord\rightarrow$}        
\def\dvec#1{\vbox{\ialign{##\crcr
        \leftrightarrowfill\crcr\noalign{\kern-1pt\nointerlineskip}
        $\hfil\displaystyle{#1}\hfil$\crcr}}}           
\def\diag{{\rm diag \,}}                                

\def\-{\hphantom{-}}

\title{\vskip -1.0in
\begin{minipage}{14.5cm}
\begin{flushright}
 {\small IFT UAM/CSIC-13-137}
\end{flushright}
\end{minipage}\\
\vskip 1.0in
Riccati equations for holographic 2-point functions  
}

\author{Ioannis Papadimitriou,${}^1$\footnote{\tt ioannis.papadimitriou@csic.es}\, and Anastasios Taliotis,${}^2$\footnote{\tt atalioti@vub.ac.be}\, \\ \\
\href{http://www.ift.uam-csic.es/en}{${}^1$}{\it Instituto de F\'{\i}sica Te\'orica UAM/CSIC} \\ {\it Universidad Aut\'onoma de Madrid}\\ {\it Madrid 28049, Spain}\\ \\
\href{http://we.vub.ac.be/tena/}{${}^2$}{\it Theoretische Natuurkunde, Vrije Universiteit Brussel}, and \\
{\it International Solvay Institutes,
Pleinlaan 2, B-1050 Brussels, Belgium}\\ \\ }

\date{}

\begin{document}

\maketitle

\begin{abstract}

Any second order homogeneous linear ordinary differential equation can be transformed into a first order non-linear Riccati equation. We argue that the Riccati form of the linearized fluctuation equations that determine the holographic 2-point functions simplifies considerably the numerical computation of such 2-point functions and of the corresponding transport coefficients by computing directly the response functions, eliminating the arbitrary source from the start. Moreover, it provides a neat criterion for the infrared regularity of the fluctuations. In particular, it is shown that the infrared regularity conditions for scalar and tensor fluctuations coincide, and hence they are either both regular or both singular. We demonstrate our numerical recipe based on the Riccati equations by computing the holographic 2-point functions for the stress tensor and a scalar operator in a number of asymptotically anti de Sitter backgrounds of bottom up scalar-gravity models. Analytical results are obtained for the 2-point function of the transverse traceless part of the stress tensor in two confining geometries, including a geometry that belongs to the class of IHQCD. We find that in this background the spin-2 spectrum is linear and, as expected, the position space 2-point function decays exponentially at large distances at a rate proportional to the confinement scale.

\end{abstract}

\newpage

\tableofcontents
\addtocontents{toc}{\protect\setcounter{tocdepth}{3}}
\renewcommand{\theequation}{\arabic{section}.\arabic{equation}}

\section{Introduction}
\label{intro}
\setcounter{equation}{0}

The AdS/CFT correspondence \cite{Maldacena:1997re} has found fruitful ground in building holographic models following top down/bottom up approaches. Such models have applications in cosmology, QCD and condensed matter  physics and often the computation of the 2-point functions \cite{Gubser:1998bc,Witten:1998qj} and the transport coefficients in the dual QFT is required.

The holographic computation of 2-point functions involves solving second order linear differential equations in an asymptotically locally anti de Sitter background. The general solution of such equations contains two arbitrary functions of the boundary coordinates or momenta, one of which is generically identified with the source of the dual gauge invariant operator while the other is identified with its 1-point function. Requiring that the solution of the linear equations be regular in the interior of the bulk imposes a relation between the two arbitrary functions that parameterize the solution, thus leaving one arbitrary function. The fact that the desired solution of the second order linear equations contains an arbitrary function renders the numerical computation of the holographic 2-point functions tedious and inefficient. However, homogeneous second order linear differential equations can always be transformed to a first order non-linear Riccati equation. Namely, starting from the second order linear equation
\be
\a(x)y''+\b(x)y'+\g(x)y=0,
\ee 
and defining 
\be
w=-\frac{y'}{\a y},
\ee
one obtains the Riccati equation
\be
w'=\a(x) w^2+\frac{\a'(x)-\b(x)}{\a(x)}w+\frac{\g(x)}{\a^2(x)}.
\ee
The general solution of the Riccati equation contains only one arbitrary constant, which can be eliminated by imposing regularity of the linearized fluctuations in the bulk. The resulting solution contains no arbitrary functions and is therefore much better suited for numerical analysis. In effect the Riccati equation computes directly the kernel of the second order linear differential equation.    

In order to demonstrate the holographic computation via the Riccati equation we consider linear fluctuations around asymptotically AdS Poincar\'e domain walls in a simple model of a neutral scalar coupled to Einstein-Hilbert gravity. Such domain walls correspond to renormalization group flows of the dual CFT, triggered by a single trace deformation or by a VEV of the relevant gauge-invariant operator dual to the bulk scalar. Our approach is completely bottom up, and so we will make no attempt to embed the scalar potential in some gauged supergravity here. Of course, our techniques and analysis apply equally well to embeddable models. 

Moreover, we will allow the domain wall backgrounds to contain mild singularities. In particular we will allow for ``good singularities'' according to \cite{Gubser:2000nd}, since these can be in principle resolved by stringy effects and/or by a horizon. In fact, such singular backgrounds have been extensively used as toy holographic models for confinement (see e.g. \cite{Gursoy:2007er} and references therein.)       

The paper is organized as follows. In section \ref{model} we present the model we are going to use as well as the Poincar\'e domain wall backgrounds in a form that will be most suitable for our subsequent analysis. In section \ref{lin-eqs} we derive the linearized fluctuation equations in both second order and first order Riccati forms, closely following the analysis of \cite{Papadimitriou:2004rz}. Section \ref{hr} deals with the ultraviolet divergences of the 2-point functions, again using the techniques of \cite{Papadimitriou:2004rz}, while in section \ref{class} we classify confining infrared geometries that allow for non-singular 2-point functions. In particular, we provide a general criterion for the infrared regularity of scalar and tensor fluctuations and in lemma \ref{lemma} we show that the assumptions of the holographic $c$-theorem ensure that such fluctuations are either both regular or both singular at the infrared. In section \ref{nure} we describe our numerical strategy while in section \ref{examples} we provide indicative exact and numerical results for three different backgrounds, namely exact AdS and two asymptotically AdS geometries that are confining in the infrared. We end with some discussion and a summary of our results in section \ref{discussion}. Some technical results are presented in the appendices.

\section{RG flows and models of confinement in AdS/CFT}
\label{model}
\setcounter{equation}{0}

In order to demonstrate the use of the Riccati equation for the computation of holographic 2-point functions we consider a bottom up gravity-scalar action 
\begin{equation}
\mathcal{S}=\frac{1}{2\kappa^{2}}\int d^{d+1}x\sqrt{-g}\left(R-\frac{1}{2}(\partial\f)^{2}-V(\f)\right),
\end{equation}
where $\k^2=8\p G_{d+1}$ is the gravitational constant in $d+1$ dimensions and the potential is left unspecified at this point, except for the requirement that the equations of motion admit AdS$_{d+1}$ as a solution. The equations of motion take the form
\bea
&&R_{\mu\nu}-\frac12 R g_{\mu\nu}=\frac{1}{2}\left(\partial_{\mu}\f\partial_{\nu}\f-\frac{1}{2}g_{\mu\nu}(\partial\f)^{2}\right)-\frac{1}{2}g_{\mu\nu}V(\f)\equiv T_{\mu\nu},\NO\\
&&\square_g\f-V_\f(\f)=0,
\eea
where the subscript $\f$ denotes derivative with respect to $\f$. 

We will compute the holographic 2-point functions in backgrounds that preserve Poincar\'e invariance in $d$ dimensions, namely
\be\label{dw}
ds^2_B=dr^2+e^{2A(r)}\h_{ij}dx^i dx^j,\quad \f=\f_B(r),
\ee
where\footnote{Occasionally we will consider Euclidean signature instead below.} $\h=\diag(-,+...+)$ is the Minkowski metric in $d$ dimensions. Such backgrounds describe renormalization group flows of the dual CFT. They have also been considered as toy models for the holographic study of QCD \cite{Gursoy:2007er,Gursoy:2007cb,Gursoy:2008bu,Gursoy:2008za,Kiritsis:2011qv}, of heavy ion collisions \cite{Gubser:2008pc,Albacete:2008vs,Albacete:2009ji,Taliotis:2010pi,Aref'eva:2014sua} and thermalization \cite{Balasubramanian:2010ce,Craps:2013iaa,Aref'eva:2013wma,Kovchegov:2007pq} in the AdS/CFT context. Although the radial coordinate $r$ is most suitable for setting up the holographic dictionary, it is often useful to use instead a `conformal' radial coordinate $z$ such that   
\be\label{coco}
ds^2_B=e^{2A(z)}\left(dz^{2}-dt^{2}+d\vec{x}^{2}\right).
\ee
One can easily switch between the two radial coordinates using
\be
e^{A(z)}dz=-dr,\quad \pa_r=-e^{-A(z)}\pa_z,
\ee
and we will freely do so in the following. In particular, from now on dots denote derivatives with respect to $r$, while primes indicate derivatives with respect to $z$.      

Inserting the ansatz (\ref{dw}) in the equations of motion leads to the set of coupled equations 
\bea
&&\dot A^2-\frac{1}{2d(d-1)}\left(\dot\f_B^2-2 V(\f_B)\right)=0,\NO\\
&&\ddot A+d \dot A^2+\frac{1}{d-1}V(\f_B)=0,\NO\\
&&\ddot\f_B+d\dot A\dot\f_B-V_\f(\f_B)=0.
\eea
It is well known that these equations are automatically solved provided a function $W(\f)$ can be found such that 
\bea\label{flow_eqs}
&&\dot A=-\frac{1}{2(d-1)}W(\f_B),\NO\\
&&\dot\f_B=W_\f(\f_B),
\eea
and $W(\f)$ is related to the scalar potential as
\be\label{VW}
V(\f_B)=\frac12\left(W_\f^2-\frac{d}{2(d-1)}W^2\right).
\ee
These relations follow most naturally from a Hamilton-Jacobi analysis of the domain wall backgrounds (\ref{dw}) \cite{Skenderis:1999mm,deBoer:1999xf,Papadimitriou:2004ap,Papadimitriou:2004rz}. 

The domain wall solutions we are interested in here are asymptotically AdS, which means that as $r\to\infty$,
$A(r)\sim r/L$, where $L$ is the AdS radius. Moreover, the weaker energy condition requires that $\ddot A \leq 0$ and so $\dot A$ is monotonically increasing along the RG flow, starting with its minimum value, $1/L$, at the far UV. This leads to the holographic $c$-theorem \cite{Freedman:1999gp}. As we approach the IR there are three mutually exclusive possibilities \cite{Gursoy:2007er}. Namely, either another AdS of a different radius is reached or a curvature singularity is found at a finite value $r_o$ of the radial coordinate or at $r\to -\infty$.   
The curvature singularity can be good or bad according to the criteria of \cite{Gubser:2000nd}. Moreover, some singularities give rise to confinement according to the Wilson loop test \cite{Gursoy:2007er,Maldacena:1998im,Kinar:1998vq,Albacete:2008dz}. Two of the examples
we will consider below belong to this class. 

The requirement that the equations of motion admit AdS of radius $L$ as a solution implies that the scalar potential takes the form\footnote{One can always redefine the scalar such that AdS of radius $L$ corresponds to $\f=0$.} 
\be\label{V-asymptotics}
V(\f)=-\frac{d(d-1)}{L^2}+\frac12 m^2\f^2+\cdots,
\ee
as $\f\to 0$, where $m$ is the mass of the scalar field and the dots stand for higher powers of the scalar field. Stability with respect to scalar perturbations requires that the mass satisfies the Breitenlohner-Freedman bound  $m^2L^2\geq -(d/2)^2$ \cite{Breitenlohner:1982jf}. We will assume additionally that $m^2<0$ so that $\f\to 0$ in the UV and the scalar field is dual to a relevant operator.    

Since the scalar potential is a priori arbitrary in our bottom up model except for the asymptotic behavior (\ref{V-asymptotics}), we can parameterize the backgrounds (\ref{dw}) in terms of the `superpotential' $W(\f_B)$. The warp factor $A(r)$, the scalar $\f_B(r)$ and the potential $V(\f)$ can then be obtained via (\ref{flow_eqs}) and (\ref{VW}). The asymptotic form (\ref{V-asymptotics}) of the scalar potential implies that $W(\f_B)$ asymptotically takes the form 
\be\label{W-asymptotics}
 W(\f_B)=\left\{\begin{matrix}
 -\frac{2(d-1)}{L}-\frac{1}{2L}(d-\D)\f_B^2+\cdots,\\
 -\frac{2(d-1)}{L}-\frac{1}{2L}\D\f_B^2+\cdots,
 \end{matrix}\right.
\ee
where the scaling dimension $\D$ of the dual scalar operator is related to the scalar mass as $m^2L^2=\D(d-\D)$. Assuming $\D> d/2$, if the superpotential behaves asymptotically as in the first line of (\ref{W-asymptotics}) then the background describes an RG flow due to a single-trace deformation of the dual CFT by a relevant operator of dimension $\D$, while if $W(\f_B)$ asymptotes to the second line of (\ref{W-asymptotics}) then the background describes an RG flow due to a vacuum expectation value of such an operator \cite{Papadimitriou:2004rz}. For $\D=d/2$ the corresponding asymptotic forms of the superpotential are \cite{Papadimitriou:2007sj} 
\be\label{W-asymptotics-BF}
 W(\f_B)=\left\{\begin{matrix}
 -\frac{2(d-1)}{L}-\frac{d}{4L}\f_B^2\left(1+\frac{1}{\log\f_B}\right)+\cdots,\\
 -\frac{2(d-1)}{L}-\frac{d}{4L}\f_B^2+\cdots.
 \end{matrix}\right.
\ee

The backgrounds (\ref{dw}) with generic scalar potential can alternatively be parameterized by specifying the wrap factor $A$. This parameterization is particularly useful for studying the infrared behavior of the background and of linear perturbations around it, as well as for numerical calculations. All other quantities can be easily obtained from the warp factor through the following relations in conformal coordinates:  
\bea\label{ficon}
&&\f_B'=\pm\sqrt{-2(d-1)\left(A''-A'^2\right)},\NO\\
&&W(\f_B)=2(d-1)e^{-A}A',\NO\\
&&V(\f_B)=-(d-1)e^{-2A}\left(A''+(d-1)A'^2\right).
\eea

\section{Linearized fluctuation equations in Riccati form}
\label{lin-eqs}
\setcounter{equation}{0}

The 2-point functions of the stress tensor and of the scalar operator dual to $\f$ can be computed by solving the linearized fluctuation equations around the backgrounds considered in section \ref{model}. In this section we provide a general derivation of these fluctuation equations in the Riccati form closely following the analysis of \cite{Papadimitriou:2004rz}.      

Without loss of generality we will consider linearized metric fluctuations that preserve the partial gauge-fixing of the metric 
\be\label{gf-metric}
ds^2=dr^2+\g_{ij}(r,x)dx^idx^j,
\ee
where $\g_{ij}$ is the induced metric on the constant $r$ hypersurfaces. In this gauge the radial canonical momenta $\p^{ij}$ and $\p_\f$ are given by 
\bea\label{momenta}
&&\p^{ij}=\frac{1}{2\k^2}\sqrt{-\g}\left(K\g^{ij}-K^{ij}\right)=\frac{\d\cs}{\d\g_{ij}},\NO\\
&&\p_\f=-\frac{1}{2\k^2}\sqrt{-\g}\dot\f=\frac{\d\cs}{\d\f},
\eea
where $K_{ij}=\frac12 \dot\g_{ij}$ is the extrinsic curvature of $\g_{ij}$ and $\cs[\g,\f]$ is the on-shell action as a function of the induced fields $\g_{ij}$ and $\f$. The AdS/CFT dictionary relates the on-shell action to the generating function of connected correlation functions and the radial canonical momenta to the corresponding 1-point functions. For our present analysis the relations (\ref{momenta}) suffice, but for a general radial Hamiltonian analysis in the context of holographic renormalization we refer the reader to \cite{Papadimitriou:2004ap,Papadimitriou:2004rz} and \cite{deBoer:1999xf,Martelli:2002sp,Bianchi:2003ug,Kanitscheider:2008kd,Haack:2010zz,McFadden:2010vh,Easther:2011wh} for related work. 

The most general linear fluctuations around the backgrounds described in section \ref{model} that preserve the partially gauge-fixed metric (\ref{gf-metric}) are of the form 
\be
\g_{ij}=\g_{ij}^B(r)+h_{ij}(r,x)=e^{2A(r)}\h_{ij}+h_{ij}(r,x),\quad \f=\f_B(r)+\vf(r,x).
\ee
The extrinsic curvature can be expressed to linear order in the fluctuations as 
\be
K^i_j=\frac12\g^{ik}\dot\g_{kj}=\dot{A}\d^i_j+\frac12\dot{S}^i_j,
\ee
where $S^i_j\equiv \g^{ik}_Bh_{kj}$ can be decomposed into irreducible components as
\be
S^i_j=e^i_j+\partial^i\e_j+\partial_j\e^i+\frac{d}{d-1}\left(\frac{1}{d}\d^i_j-\frac{\partial^i\partial_j}{\square_B}\right)f+
\frac{\partial^i\partial_j}{\square_B}S,
\ee
with $\partial_i e^i_j=e^i_i=\partial_i\e^i=0$ and indices are raised with the inverse background metric $e^{-2A}\h^{ij}$. Conversely, all irreducible components can be expressed in terms of $S^i_j$ as
\be
e^i_j=\P^i\phantom{}_k\phantom{}^l\phantom{}_j S^k_l,\,\,\,\,
\e_i=\cp^l_i\frac{\partial_k}{\square_B}S^k_l,\,\,\,\,f=\cp^l_k S^k_l,\,\,\,\,
S=\d^l_k S^k_l,
\ee
via the projection operators
\be
\P^i\phantom{}_k\phantom{}^l\phantom{}_j=\frac12\left(\cp^i_k\cp^l_j+\cp^{il}\cp_{kj}-\frac{2}{d-1}\cp^i_j\cp^l_k\right),
\ee
and
\be
\cp^i_j=\d^i_j-\frac{\partial^i\partial_j}{\square_B}.
\ee
However, diffeomorphism invariance in the transverse directions can be used to set $\e_i\equiv 0$. 

Inserting these expressions for the extrinsic curvature and for the fluctuations in the equations of motion for the gauge-fixed metric (\ref{gf-metric}) (Gauss-Codazzi equations) leads to a set of linear equations for the fluctuations \cite{Papadimitriou:2004rz}
\bea\label{fluctuation-eqs}
&&\left(\partial_r^2+d\dot{A}\partial_r+e^{-2A}\square\right) e^i_j=0,\NO\\
&&\left(\partial_r^2+[d\dot{A}+2W\partial^2_\f\log W]\partial_r+e^{-2A}\square\right)\x=0,\NO\\
&&\dot{f}=-W_\f\vf,\NO\\
&&\dot{S}=-\frac{2}{W}\left[-e^{-2A}\square f+\left(W_\f\dot{\vf}-V_\f\vf\right)\right],
\eea
where $\square_B=e^{-2A}\square=e^{-2A}\h^{ij}\partial_i\partial_j$ and\footnote{Note that $\x$ is not well defined when $\f_B\equiv 0$, i.e. when the background is exactly AdS. However, we will consider this case separately in one of our examples below.}
\be
\x\equiv \frac{W}{W_\f}\vf+f.
\ee 
The modes $e^i_j$ and $\x$ satisfy decoupled second order linear equations. The solution of these second order equations captures all the non-trivial physics contained in the 2-point functions of the stress tensor and of the scalar operator $\co(x)$ dual to $\f$. However, these linear second order equations can be transformed into first order non-linear equations of Riccati type, which are much more amenable to numerical analysis.   

To derive the Riccati form of these equations we note that to linear order in the fluctuations we can write
\be\label{response}
\dot{e}^i_j=E(A,\f_B)e^i_j,\,\,\,\,\,\dot{\x}=\Om(A,\f_B)\x,
\ee
where $E(A,\f_B)$ and $\Om(A,\f_B)$ are the {\em response functions} and depend  only on the background. Inserting these expressions in the second order fluctuation equations for $e^i_j$ and $\x$ leads to {\em first order} Riccati equations for the response functions $E(A,\f_B)$ and $\Om(A,\f_B)$, which in momentum space ($p^2=-\om^2+\vec k^2$) are \cite{Papadimitriou:2004rz}
\begin{align}
\label{master}
\boxed{
\begin{aligned}
&\dot{E}+E^2+d\dot{A}E-e^{-2A}p^2=0, \\
&\dot{\Om}+\Om^2+\left(d\dot{A}+2W\partial^2_\f\log W\right)\Om-e^{-2A}p^2=0.
\end{aligned}}
\end{align}
Contrary to the second order equations for the modes $e^i_j$ and $w$, these first order equations only require one boundary condition each. As we will discuss in section \ref{class}, this boundary condition is provided by imposing regularity of the fluctuations $e^i_j$ and $w$ in the infrared. The fact that the Riccati equations (\ref{master}) directly compute the response functions bypassing the arbitrary sources provides a much more efficient strategy for computing the 2-point functions numerically. Moreover, as we now show, these 2-point functions can be directly expressed in terms of the response functions $E(A,\f_B)$ and $\Om(A,\f_B)$ so that the 2-point functions can be simply red off from the solutions of (\ref{master}).  

The connection between the response functions $E(A,\f_B)$ and $\Om(A,\f_B)$ and the 2-point functions can be easily shown by using the canonical momenta (\ref{momenta}). Namely, given the response functions the radial velocities can be written as
\bea
&&\dot{e}^i_j=Ee^i_j,\NO\\
&&\dot{f}=-W_\f\vf,\NO\\
&&\dot{S}=-2\left[\left(\frac{W_\f}{W}\right)^2\Om-\frac{e^{-2A}}{W}\square\right]f-2\frac{W_\f}{W}\left(\Om+\frac{d}{2(d-1)}W\right)\vf,\NO\\
&&\dot{\vf}=(W_{\f\f}+\Om)\vf+\frac{W_\f}{W}\Om f.
\eea
Moreover, expanding the canonical momenta (\ref{momenta}) to linear order in the fluctuations we get
\bea
&&\p^{ij}=\frac{1}{2\k^2}\sqrt{-\g_B}\left((d-1)\dot A\g_{B}^{ij}-(d-1)\dot A\left(S^{ij}-\frac12\g_B^{ij}S\right)
-\frac12\left(\g_B^{(ik}\dot S^{kj)}-\g_B^{ij} \dot S\right)\right),\NO\\
&&\p_\f=-\frac{1}{2\k^2}\sqrt{-\g_B}\left(\dot\f_B+\dot\vf+\frac12\dot\f_B S\right).
\eea
Inserting the radial velocities in these expansions of the momenta and isolating the terms linear in the fluctuations gives
\bea
\p^{(2)ij}&=&\frac{1}{4\k^2}\sqrt{-\g_B}\left\{(W-E)e^{ij}-W\left(\cp^{ij}-\frac12\g_B^{ij}\right)S\right.\NO\\
&&\left.+W\left(\frac{d}{d-1}\cp^{ij}-\g_B^{ij}\right)f-2\cp^{ij}\left[\left(\frac{W_\f}{W}\right)^2\Om-\frac{e^{-2A}}{W}\square\right]f\right.\NO\\
&&\left.
-W_\f\left(\frac{2\Om}{W}\cp^{ij}+\g_B^{ij}\right)\vf\right\},
\eea
and
\be
\p_\f^{(2)}=-\frac{1}{2\k^2}\sqrt{-\g_B}\left((W_{\f\f}+\Om)\vf+\frac{W_\f}{W}\Om f+\frac12 W_\f S\right).
\ee
According to (\ref{momenta}) these canonical momenta can be expressed as gradients of the quadratic in fluctuations part of the on-shell action, which we will denote by $\cs^{(2)}$. Indeed, these expressions for the momenta can be integrated straightforwardly to obtain the generating functional for all 2-point functions, namely
\begin{align}
\label{2-point-fns}
\begin{aligned}
&\cs^{(2)}=-\frac{1}{8\k^2}\int d^dx \sqrt{-\g_B}\left\{(E-W)e^i_je^j_i+2W f S+\frac12 W S^2 +2W_\f\left(\frac{2\Om}{W}f+S\right)\vf\right.\\
&\left.+2f\left(-\frac{dW}{2(d-1)}+\left(\frac{W_\f}{W}\right)^2\Om-\frac{e^{-2A}}{W}\square\right)f
+2\vf(W_{\f\f}+\Om)\vf\right\}.
\end{aligned} 
\end{align}
This expression encodes all 2-point functions between the stress tensor and the operator $\co(x)$, but it suffers from ultraviolet divergences which must be consistently removed. Moreover, many terms in (\ref{2-point-fns}) are trivial contact terms that can be removed by finite local counterterms. We address both these issues in the subsequent section.

\section{Renormalized 2-point functions}
\label{hr}
\setcounter{equation}{0}

In order to obtain the renormalized 2-point functions from the generating functional (\ref{2-point-fns}) we need to add local covariant boundary terms to remove the ultraviolet divergences. Moreover, we are free to add any finite local counterterms we find convenient, which reflects the usual renormalization scheme choice. This freedom can be utilized to greatly simplify (\ref{2-point-fns}). Namely, we observe that all terms that don't contain $E$ and $\Om$ 
in (\ref{2-point-fns}) are local covariant terms that can be simply removed by local counterterms.\footnote{More correctly, here we are ignoring for simplicity all non-analytic terms in the 2-point functions. In the case of backgrounds describing RG flows due to a vacuum expectation value there are generically contact terms that are proportional to the VEVs and these should not be removed by local counterterms as they are physical. But such terms can be recovered easily by a careful analysis of the counterterms. We refer the reader to \cite{Papadimitriou:2004rz} for a more complete analysis of this issue.} The generating functional (\ref{2-point-fns}) can therefore be simplified to 
\begin{align}
\label{2-point-fns-ct1}
\begin{aligned}
&\cs^{(2)}=-\frac{1}{8\k^2}\int d^dx \sqrt{-\g_B}\left\{e^j_iEe^i_j+2\left(\vf+\frac{W_\f}{W}f\right)\Om \left(\vf+\frac{W_\f}{W}f\right)\right\}.
\end{aligned} 
\end{align}
Note that now only the decoupled dynamical modes $e^i_j$ and $w$ survive after trivial contact terms are removed. 

However, this simplified generating function still suffers from ultraviolet divergences. The local covariant counterterms that are required to remove these divergences depend on the particular theory, i.e. the scalar potential $V(\f)$, and can be systematically obtained directly at the linearized level by the algorithm described in 
\cite{Papadimitriou:2004rz}. Namely, the response functions $E$ and $\Om$ can be expanded in eigenfunctions of the dilatation operator 
\be
\label{dilatation-op}
\d_D=\pa_A+(\D-d)\f_B\pa_{\f_B},
\ee
as 
\bea\label{UV-expansions}
&&E=E\sub{1}+\cdots+\wt E\sub{d}\log(e^{-2r/L})+E\sub{d}+\cdots,\NO\\
&&\Om=\left\{\begin{matrix}
&\Om\sub{1}+\cdots+\wt \Om\sub{2\D-d}\log(e^{-2r/L})+\Om\sub{2\D-d}+\cdots, & \D> d/2,\\
&\frac{L}{r}\wt \Om\sub{0}+\frac{L^2}{r^2}\Om\sub{0}+\cdots, & \D=d/2,
\end{matrix}\right.
\eea
where the subscript indicates the eigenvalue under the dilation operator, e.g. $\d_D E\sub{k}=-k E\sub{k}$. The terms up to $\wt E\sub{d}$ and $\wt \Om\sub{2\D-d}$ are local, contribute to the ultraviolet divergences, and can be determined by inserting these expansions in the equations (\ref{master}) for the response functions and using the fact that 
\be
\pa_r=\dot A\pa_A+\dot\f_B\pa_{\f_B}=-\frac{1}{2(d-1)}W(\f_B)\pa_A+W_\f(\f_B)\pa_{\f_B}\sim \d_D+\cdots. 
\ee
These terms must be removed from the generating functional by adding the corresponding local counterterms. The terms $E\sub{d}:=e^{-dr/L}\Hat E\sub{d}$ and $\Om\sub{2\D-d}:=e^{-(2\D-d)r/L}\Hat\Om\sub{2\D-d}$ however are in general non-local and they are left undetermined by this asymptotic analysis of (\ref{master}). $\Hat E\sub{d}$ and $\Hat\Om\sub{2\D-d}$ defined so that they are independent of the radial coordinate and correspond to the renormalized response functions that contain all the physical information of the renormalized 2-point functions. Terms of higher order than $E\sub{d}$ and $\Om\sub{2\D-d}$ drop out when the UV cutoff is removed and so we need not consider them. We will see explicit examples of these expansions in section \ref{examples}. 

The final outcome of this analysis is that the renormalized generating functional for the 2-point functions can be written in the form  
\begin{align}
\label{2-point-fns-ren}
\boxed{
\begin{aligned}
&\cs^{(2)}_{ren}=-\frac{1}{8\k^2}\lim_{r\to\infty}\int d^dx \sqrt{-\g_B}\left\{e^j_iE\sub{d}e^i_j+2\left(\vf+\frac{W_\f}{W}f\right)\Om\sub{2\D-d} \left(\vf+\frac{W_\f}{W}f\right)\right\}.
\end{aligned} }
\end{align}
All renormalized 2-point functions can be simply red off this expression. Namely, 
\begin{align}
\label{ren-2pt-fns}
\boxed{
\begin{aligned}
&\langle T_{ij}(p)T_{kl}(-p)\rangle_{TT}=-\frac{1}{\k^2}\P_{ijkl}\Hat E\sub{d}(p),\\ 
&\langle \co(p)\co(-p)\rangle=-\frac{1}{2\k^2}\Hat\Om\sub{2\D-d}(p),\\
&\langle \cp^{ij}T_{ij}(p)\co(-p)\rangle=\frac{1}{2\k^2}\frac{(d-\D)}{(d-1)}\Hat\f_B\Hat\Om\sub{2\D-d}(p),\\
&\langle \cp^{ij}T_{ij}(p)\cp^{kl}T_{kl}(-p)\rangle=-\frac{1}{2\k^2}\frac{(d-\D)^2}{(d-1)^2}\Hat\f_B^2\Hat\Om\sub{2\D-d}(p),
\end{aligned} }
\end{align}
where $\Hat\f_B$ is the scalar source of the background defined through the asymptotic relation $\f_B\sim z^{d-\D}\Hat\f_B$. Note that if the background domain wall describes an RG flow due to a VEV of the operator $\co$, then the source $\Hat\f_B$ vanishes and so there is no mixing between the stress tensor and the scalar operator.

These expressions for the renormalized 2-point functions reduce the problem to the evaluation of the renormalized response functions $\Hat E\sub{d}(p)$ and $\Hat\Om\sub{2\D-d}(p)$. To compute these one simply needs to solve the Riccati equations (\ref{master}) exactly, analytically or numerically, and to identify the renormalized response functions by subtracting the terms in the asymptotic expansions (\ref{UV-expansions}). A number of examples will be worked out explicitly in section \ref{examples}.

\section{Infrared regularity conditions}
\label{class}
\setcounter{equation}{0}

In this section we classify various IR behaviors of the warp factor $A(z)$ and we determine the corresponding IR behavior of the response functions $E$ and $\Om$ following from (\ref{master}). The definitions (\ref{response}) of the response functions imply that the linear fluctuations $e^i_j$ and $w$ are regular in the IR provided
\begin{align}
\label{IRreg}
\boxed{
\begin{aligned}
&\int^r dr' E(r',p)=-\int^{z}dz' e^{A(z')}E(z',p) <\infty,\\
&\int ^r dr' \Omega(r',p)=-\int^{z} dz' e^{A(z')}\Omega(z',p) <\infty.
\end{aligned} }
\end{align}
We will now determine the appropriate IR asymptotic solutions of $E$ and $\Om$ for various choices of warp factors such that these conditions are satisfied. Noting that the quantity $W\partial^2_\f\log W$ can be expressed in terms of the warp factor as 
\be
W\pa^2_\f\log W = -\frac12e^{-A}\pa_z\log\left(1-\frac{A''}{A'^2}\right),
\ee 
the equations (\ref{master}) in Poincar\'e coordinates take the form\footnote{The transformation 
$\Om=\wt\Om \left(1-\frac{A''}{A'^2}\right)^{\frac{1}{d-1}}$ renders the second equation in (\ref{masterz}) in the same form as the first one, but with $A(z)$ replaced by the quantity $A+\frac{1}{d-1}\log\left(1-\frac{A''}{A'^2}\right)$.}
\begin{align}
\label{masterz}
\boxed{
\begin{aligned}
&E'-e^{A}E^2+d A' E +p^2 e^{-A}=0, \\
&\Om'-e^A \Om^2+\left(dA'+\pa_z\log\left(1-\frac{A''}{A'^2}\right)\right)\Om+e^{-A}p^2=0,
\end{aligned} }
\end{align}
where the primes denote differentiation w.r.t. $z$. This form of the equations for the response functions allows us to prove the following 
\begin{lemma}\label{lemma}
For any asymptotically AdS Poincar\'e domain wall that fulfills the conditions of the holographic $c$-theorem \cite{Freedman:1999gp}, the asymptotic behaviors of the response functions $E$ and $\Om$ in the IR are identical provided the IR geometry is not another AdS, in which case the answer depends on how fast AdS is approached. 
\end{lemma}
The coefficient of $\Om$ in (\ref{masterz}) can be written as
\be
\pa_z\log\left[\left(1-\frac{A''}{A'^2}\right)e^{dA}\right].
\ee
Hence, given that $A\to -\infty$ in the IR, and excluding the case that the IR geometry is another AdS, i.e. $A(z)\sim -\log z$ as $z\to\infty$, it suffices to show that $A''/A'^2$ does not approach $-\infty$ in the IR. The holographic $c$-theorem requires that $\ddot A\leq 0$, or $A''\leq A'^2$. Hence, $\dot A$ is monotonically decreasing towards the UV. Since at the far UV $\dot A \to 1/L>0$, it follows that $\dot A >0$ along the entire flow, or equivalently $A'\leq 0$.
Since $-1/A'\geq 0$ is bounded from below, $\pa_z(-1/A')$ can only possibly tend to $-\infty$ at a finite value $z_0$ of $z$ and not asymptotically as $z\to\infty$. In that case $-1/A'$ has a branch cut at $z=z_0$ such that 
\be
-\frac{1}{A'}\sim a+b(z_0-z)^\a,\quad a\geq0,\quad b>0,\quad 0<\a<1,
\ee  
as $z\to z_0^-$. Hence, 
\be
A(z)\sim \left\{\begin{matrix}
\frac{1}{a}(z_0-z)+const. & a>0, \\ 
\frac{1}{(1-\a)b}(z_0-z)^{1-\a}+const. & a=0, 
\end{matrix}\right.
\ee
which contradicts the hypothesis that $A(z)\to -\infty$ at the IR. If $A(z)\sim -\log z$ as $z\to\infty$, then $1-A''/A'^2\to 0$ and so we need subleading terms to determine which of the two terms multiplying $\Om$ in (\ref{masterz}) falls off faster. $\Box$ 

This lemma implies that it suffices to examine the IR behavior of $E$ only, since that of $\Om$ is identical. Moreover, we will work in Euclidean signature in this section so that $p^2\geq 0$.  

\begin{enumerate}

\item $A(z)\sim - c \log z,\quad c>0,$ as $z\to z_{IR}=+\infty$:  

The equation for $E$ in (\ref{master}) in the IR becomes 
\be
E'-z^{-c} E^2-dc z^{-1} E+p^2 z^c \approx0.
\ee
Trying a solution  $E\sim E_o z^\e$ we get 
\be
\e E_o z^{\e-1}-E_o^2 z^{2\e-c}-dc E_o z^{\e-1}+p^2 z^c\approx 0. 
\ee
The only solutions are 
\be
\e=c,\quad E_o=\pm \sqrt{p^2}. 
\ee
The positive solution for $E_o$ satisfies the regularity condition (\ref{IRreg}) for the fluctuations and so there exists a regular mode.

\item $A(z)\sim -c z^\a,\quad c,\a > 0$,  as $z\to z_{IR}=+\infty$:

Taking $E\sim E_o e^{c z^\a}$ we have 
\be
-(d-1)c\a z^{\a-1} E_o + (p^2-E_o^2)\sim 0. 
\ee
Moreover, the regularity condition requires only that $E_o>0$. We have three cases to distinguish. 

\begin{enumerate}

\item For $\a<1$ we have $E_o=\sqrt{p^2}$, which as in the previous example leads to a regular solution at least in Euclidean space. In Lorentzian space, according to \cite{Gursoy:2007er}, such a geometry is not confining independently of the coefficient $c$.

\item For $\a=1$ the two solutions for $E_o$ are 
\be
E_o=-\frac{(d-1)c}{2}\pm\sqrt{\left(\frac{(d-1)c}{2}\right)^2+p^2}.
\ee  
According to \cite{Gursoy:2007er}, the corresponding geometry  yields a mixed quantized and continuum spectrum.
In the Euclidean case, only the coefficient with the positive sign leads to a regular solution. 
 
\item Finally, for $\a>1$ there is no regular solution with the above ansatz for $E$ but there is one for the slightly more general ansatz $E\sim E_o z^a e^{c z^\a}$. Namely, the choice $a=1-\a$ and 
\be
E_o=\frac{p^2}{(d-1)c\a},
\ee
leads to regular fluctuations according to (\ref{IRreg}). This background is confining according to the Wilson loop test \cite{Gursoy:2007er}.

\end{enumerate}

In all, for every $\a>0$ there exists a regular solution.

\item $A(z)\sim  c \log (z_0-z),\quad c>0,$ as $z\to z_{IR}=z_0$:  

Using the ansatz $E\sim E_o (z_0-z)^a e^{-A(z)}$ we have 
\be
-(a+(d-1)c) E_o(z_0-z)^{a-1}-E_0^2(z_0-z)^{2a} + p^2\approx 0.
\ee
The choice $a=1$ and $E_o =\frac{ p^2}{(d-1)c+a}$ leads to regular fluctuations according to (\ref{IRreg}).

\item $A(z)\sim  -c(z_0-z)^{-a},\quad c>0,\,\, a>0$ as $z\to z_{IR}=z_0$:  

Taking the ansatz $E\sim E_o (z_0-z)^k e^{-A(z)}$ we have 
\be
-(d-1)acE_o(z_0-z)^{k-1-a}-kE_o(z_0-z)^{k-1}-E_o^2(z_0-z)^{2k}+p^2\approx 0.
\ee
This admits a solution $k=1+a$ and
\be
E_o = \frac{p^2}{a c (d-1)}, 
\ee
which leads to regular fluctuations according to (\ref{IRreg}). 

\end{enumerate}

\section{Numerical recipe}
\label{nure}
\setcounter{equation}{0}

In this section we outline a recipe for the numerical evaluation of the renormalized response functions $\Hat E\sub{d}$ and $\Hat\Om\sub{2\D-d}$ that are required to obtain all the 2-point functions of the stress tensor and the scalar operator $\co$. The formulation of the 2-point functions in  terms of the Riccati equations (\ref{master}) for the response functions allows us to develop a much more efficient algorithm than the usual one based on the second order equations obeyed by the fluctuations.

\vspace{0.2in}
The recipe involves the following steps.

\begin{enumerate}

\item  {\bf Background:} 

Specification of the background by providing the warp factor $A(z)$ in conformal coordinates. All other quantities of the background can be deduced from the warp factor via (\ref{ficon}).

\item {\bf Dimensionless parameters:}

The response functions $E$ and $\Omega$ are determined by the first order equations (\ref{masterz}). Since $E$ and $\Omega$ have dimensions of mass, it is convenient to introduce the dimensionless quantities $\mathbf{E}:=L E$ and $\mathbf{\Om}:=L \Om$. Moreover, for any background other than exact $AdS$ there is a dynamical mass scale, $\L$, that governs the IR behavior of the background. This allows us to introduce the dimensionless variable $w:=\L z$. The warp factor can then be written as a function $A(w;L\L)$ of the dimensionless variable $w$ and the dimensionless parameter $L\L$. Generically the equations (\ref{masterz}), when expressed in terms of dimensionless variables, will depend on the dimensionless parameters $p/\L$ and $L\L$ and so $\mathbf{E}(w;L\L,p/\L)$ and $\mathbf{\Om}(w;L\L,p/\L)$ will be functions of these parameters too. In this most general case the renormalized response functions must be determined as functions of the two dimensionless parameters $p/\L$ and $L\L$. In the examples we will consider in section \ref{examples}, however, the equations (\ref{masterz}), when expressed in terms of dimensionless variables, only depend on the dimensionless parameter $p/\L$ and not $L\L$. As a result $\mathbf{E}(w;p/\L)$ and $\mathbf{\Om}(w;p/\L)$ are only functions of the dimensionless parameter $p/\L$.
In order to determine the renormalized response functions in this case we need to solve (\ref{masterz}) numerically for various values of the dimensionless parameter $p/\L$, and then extract the appropriate coefficient as a function of $p/\L$. Since there is only one dimensionless parameter in the examples of section \ref{examples} we can simplify things further by setting $\L=1$.

\item  {\bf IR solution:}

In order to solve (\ref{masterz}) for the response functions we must fix the single integration constant for each equation by imposing regularity of the corresponding fluctuations in the IR, i.e. imposing the conditions (\ref{IRreg}). This is done by constructing the asymptotic solutions of (\ref{masterz}) and identifying the ones that satisfy the regularity conditions (\ref{IRreg}).

\item {\bf IR shooting:}

The regular IR asymptotic solutions of the previous step are now used as IR boundary conditions in order to solve
(\ref{masterz}) numerically via a shooting procedure. In particular, one chooses a starting point $w_{in}$ very close to the IR where the numerical solution is matched with the asymptotic IR solution. The resulting numerical solution matches the desired IR expansion very accurately in the far IR, while the two solutions start to deviate away from the IR region (e.g. left panel in figs. \ref{OmIId4} and \ref{OmIId3} and fig. \ref{gIRfig}).

\item {\bf Extracting UV data:} 

In the last step the renormalized response functions $\Hat E\sub{d}$ and $\Hat\Om\sub{2\D-d}$ are extracted from the numerical solution obtained in the previous step. This is achieved by comparing the numerical solution with the general UV asymptotic solutions of (\ref{masterz}), parameterized in terms of $\Hat E\sub{d}$ and $\Hat\Om\sub{2\D-d}$. Repeating this step for various values of the dimensionless parameter $p/\L$ allows us to reconstruct the renormalized response functions as functions of the momentum, and hence compute the corresponding 2-point functions.

\end{enumerate}

\section{Examples}
\label{examples}
\setcounter{equation}{0}

In this section we consider three particular backgrounds, exact AdS and two confining geometries, in order to demonstrate the use of the Riccati equations for the computation of the 2-point functions.  

\subsection{Unbroken conformal symmetry}

Our first example is exact AdS space, i.e. 
\be 
A(z)=\log \left(\frac{L}{z}\right), 
\ee
where $L$ is the AdS radius and the boundary is at $z=0$. This implies that $\f_B=0$ and hence
\be
W(\f_B)=-\frac{2(d-1)}{L},\quad V(\f_B)=-\frac{d(d-1)}{L^2}.
\ee
The fluctuation equation for $\mathbf{E}$ (\ref{master}) takes the form
\be\label{masterEI}
z\mathbf{E}'-\mathbf{E}^2-d\mathbf{E}+p^2z^2=0,
\ee
where $\mathbf{E}:= EL$. However, the mode $\x$ is not well defined in this case since $W_\f(\f_B)=0$ and so the equation for the scalar fluctuations need to be revisited. This can be easily done by rewriting the second order equation for $\x$ in (\ref{fluctuation-eqs}) as an equation for $\wt \x := (W_\f/W) \x$ and then setting $W_\f=0$. This leads to a second order linear equation $\wt \x=\vf$, namely
\be
\ddot\vf+\frac{d}{L}\dot\vf-m^2\vf+e^{-2r/L}\square\vf=0,
\ee
where $m^2$ is the scalar mass defined in (\ref{V-asymptotics}). Writing again
\be
\dot\vf=F\vf,
\ee 
and changing to the $z$ coordinate leads to the Riccati equation 
\be\label{masterFI}
z\mathbf{F}'-\mathbf{F}^2-d\mathbf{F}+p^2z^2+m^2=0,
\ee
where $\mathbf{F}:= F L$. Note that (\ref{masterEI}) is a special case of (\ref{masterFI}) corresponding to $m^2=0$. Introducing the variable $u=p^2z^2$ (\ref{masterFI}) becomes identical to equation (3.1.24) in \cite{Papadimitriou:2010as}. The general solution takes the form
\be
\mathbf{F}_\pm=-\frac d2\pm\frac{pz\left(K'_{k}\left(pz\right)+cI'_{k}\left(pz\right)\right)}{K_{k}\left(pz\right)+cI_{k}\left(pz\right)},
\ee  
where $k=\D-d/2>0$, $c(p)$ is an integration constant, and the primes here denote differentiation with respect to the argument of the Bessel functions and not $z$. Moreover we have defined $p:=\sqrt{p^2}$, working for simplicity in Euclidean signature so that $p^2\geq 0$. It follows that 
\be
\vf=z^{d/2}\left(K_{k}\left(pz\right)+cI_{k}\left(pz\right)\right)^{\mp 1}\vf\sub{0}(p),
\ee
and hence only $\mathbf{F}_-$ leads to a solution of the second order equation for $\vf$. Requiring that $\vf$ remains finite in the IR forces us to set $c=0$ so that the desired solution for $\mathbf{F}$ is 
\be
\mathbf{F}=-\frac d2-\frac{pz K'_{k}\left(pz\right)}{K_{k}\left(pz\right)}.
\ee  
In this case the expansion of $F$ asymptotically in eigenfunctions of the dilatation operator can be simply obtained by expanding the Bessel functions for small $z$. Taking for concreteness $d$ to be even this leads to the renormalized response functions \cite{Papadimitriou:2010as} 
\be\label{responseI}\boxed{
L F\sub{2\D-d}=\frac{(-1)^{k}}{2^{2k-1}\G(k)^2}z^{2\D-d}\left(p^2\right)^{k}\log\left(\frac{p^2}{\m^2}\right),\quad 
L E\sub{d}=\frac{(-1)^{d/2}}{2^{d-1}\G(d/2)^2}z^{d}\left(p^2\right)^{d/2}\log\left(\frac{p^2}{\m^2}\right),}
\ee
where $\m$ is an arbitrary energy scale.

\subsection{A toy model for confinement}

Next we consider a warp factor of the form
\be\label{bgdII}
A(z)=\frac{2}{d-1}\log\left(1-\L^{d}z^{d}\right)+\log \left(\frac{L}{z}\right),
\ee
where $\L$ is a positive constant with the dimension of mass. This geometry is confining according to the Wilson loop test, with confinement scale set by $\Lambda$. Using the relations (\ref{ficon}) we determine\footnote{A closed form expression for $\f_B(z)$ can be obtained in terms of Appel functions.}  
\be
\f_B'^2=\frac{4d(d+1)z^{d-2}\L^d\left(d-1+\L^d z^d\right)}{(d-1)\left(1-\L^d z^d\right)^2},
\ee
as well as
\be
W(z)=-\frac{2}{L}\left(1-\L^d z^d\right)^{-\frac{d+1}{d-1}}\left(d-1+(d+1)\L^d z^d\right),
\ee
and 
\be
V(z)=-\frac{d}{L^2}\left(1-\L^{d}z^{d}\right)^{-2\left(\frac{d+1}{d-1}\right)}\left(d-1+(d+1)\L^{2d} z^{2d}\right).
\ee
Moreover, the UV asymptotic expansion of the background scalar is 
\be
\f_B(z)=\pm 4\sqrt{\frac{d+1}{d}}\left((\L z)^{d/2}+\frac{2d-1}{3d(d-1)}(\L z)^{3d/2}+\co(z^{5d/2})\right),
\ee
which determines the asymptotic form of the superpotential and the scalar potential to be
\be
W(\f_B)=-\frac{2(d-1)}{L}-\frac{1}{2L}\left(\frac d2 \right)\f_B^2+\co(\f_B^4),
\ee
and
\be
V(\f_B)=-\frac{d(d-1)}{L^2}-\frac{1}{2L^2}\left(\frac d2 \right)^2\f_B^2+\co(\f_B^4).
\ee
Comparing these with (\ref{W-asymptotics}) and (\ref{V-asymptotics}) we conclude that $\D=d/2$ and so the mass saturates the BF bound. Moreover this background describes a VEV and not a deformation. This is therefore exactly analogous to the Coulomb branch flow of $\cn=4$ super Yang-Mills \cite{Kraus:1998hv,Freedman:1999gk}, which was analyzed using the Riccati form of the fluctuation equations in \cite{Papadimitriou:2004rz}.

\begin{flushleft}
{\bf Riccati equations}
\end{flushleft}

The Riccati equations (\ref{master}) for the response functions take the form\footnote{We should point out that the limit $\L\to 0$ in the equation for $E$ gives the correct result, but it does not in the equation for $\Om$. The correct equations for $\L=0$ are those corresponding to the first example above.} 
\bea\label{masterII}
&&w\pa_w\mathbf{E}-\left(1-w^d\right)^{\frac{2}{d-1}}\mathbf{E}^2-d\left(1+\frac{2dw^d}{(d-1)(1-w^d)}\right)\mathbf{E}
+\left(1-w^d\right)^{-\frac{2}{d-1}}\frac{p^2}{\L^2}w^2=0,\NO\\
&&w\pa_w\mathbf{\Om}-\left(1-w^d\right)^{\frac{2}{d-1}}\mathbf{\Om}^2+\left(1-w^d\right)^{-\frac{2}{d-1}}\frac{p^2}{\L^2}w^2\NO\\
&&\phantom{more}+d\left(\frac{d+1}{d-1}-\frac{2d}{(d-1)(1-w^d)}-\frac{d-1}{d-1+w^d}+\frac{2(d-1)}{d-1+(d+1)w^d}\right)\mathbf{\Om}=0,
\eea
where we have introduced $w:=\L z$, and $\mathbf{E}:= EL$ and $\mathbf{\Om}:= \Om L$ as before. 

\begin{flushleft}
{\bf IR asymptotic solutions}
\end{flushleft}

The IR is located at $w=1$. The IR behavior of the response functions $E$ and $\Om$ can therefore be obtained from the form of (\ref{masterII}) in the vicinity of $w=1$, namely 
\bea
&&w\pa_w\mathbf{E}-\left(1-w^d\right)^{\frac{2}{d-1}}\mathbf{E}^2-\frac{2d^2}{(d-1)(1-w^d)}\mathbf{E}
+\left(1-w^d\right)^{-\frac{2}{d-1}}\frac{p^2}{\L^2}\approx 0.
\eea
$\bf{\Om}$ satisfies an identical in the IR, as ensured by lemma \ref{lemma}. The IR solutions that satisfy the regularity conditions (\ref{IRreg}) are
\be\label{IR-reg-II}\boxed{
\mathbf{E}\sim \mathbf{\Om}\sim \frac{p^2}{3d\L^2}\left(1-w^d\right)^{\frac{d-3}{d-1}}.}
\ee

\begin{flushleft}
{\bf UV asymptotic solutions}
\end{flushleft}

To determine the UV behavior of the response functions we can, without loss of generality, drop all terms containing $w^d$ in (\ref{masterII}), which gives
\bea
&&w\pa_w\mathbf{E}-\mathbf{E}^2-d\mathbf{E}+\frac{p^2}{\L^2}w^2\approx 0,\NO\\
&&w\pa_w\mathbf{\Om}-\mathbf{\Om}^2+\frac{p^2}{\L^2}w^2\approx 0.
\eea
The first equation is identical to (\ref{masterEI}) above and so the covariant asymptotic expansion for $E$, taking for concreteness $d$ even, is given by \cite{Papadimitriou:2004rz,Papadimitriou:2010as}
\bea\label{UV-asymptotics-EII}
\mathbf{E}&=&\frac{p^2z^2}{d-2}-\frac{(p^2)^2z^4}{(d-2)(d-4)}+\cdots 
+\frac{(-1)^{d/2}}{2^{d-1}\G(d/2)^2}z^{d}\left(p^2\right)^{d/2}\log\left(z^2\m^2\right)+\cdots,\\
&=&\frac{p^2L^2}{d-2}e^{-2A}-\frac{(p^2L^2)^2}{(d-2)(d-4)}e^{-4A}+\cdots \NO\\
&&\hskip2.0cm+\frac{(-1)^{d/2}}{2^{d-1}\G(d/2)^2}e^{-d A}\left(L^2p^2\right)^{d/2}\log\left(z^2\m^2\right)+e^{-dA}\Hat E\sub{d}(p)+\cdots,\NO
\eea
where $\m$ is an arbitrary dimensionful constant. Moreover, the asymptotic form of $\Om$ is 
\be\label{UV-asymptotics-OmII}
\mathbf{\Om}=-\frac{1}{\log z}+\frac{1}{\log^2 z}\Hat\Om\sub{0}(p)+\cdots.
\ee
For $d=4$ these asymptotic expansions are in agreement with those for the Coulomb branch flow given in \cite{Papadimitriou:2004rz}.

\begin{flushleft}
{\bf Exact solution for $\mathbf{E}$}
\end{flushleft}

The equation for $\mathbf{E}$ in (\ref{masterII}) can be solved analytically. The general solution is 
\be
\mathbf{E}=-\frac{(1-w^d)^{-\frac{2}{d-1}}w s'_c(w)}{s_c(w)},
\ee 
where 
\be
s_c(w):=\frac{w^{d/2}}{1-w^d}\left(K_{d/2}\left(pw/\L\right)+cI_{d/2}\left(pw/\L\right)\right),
\ee
and $c(p)$ is an integration constant. As in the first example above, the criterion (\ref{IRreg}) for the IR regularity of the fluctuations requires that we set $c=0$. Expanding the solution with $c=0$ in the UV and subtracting the asymptotic form (\ref{UV-asymptotics-EII}) we obtain the renormalized response function
\be\boxed{
LE\sub{d}=\frac{(-1)^{d/2}}{2^{d-1}\G(d/2)^2}z^{d}\left(p^2\right)^{d/2}\log\left(\frac{p^2}{\m^2}\right),}
\ee   
where $\m$ is a dimensionful constant. It follows that the non analytic part of the 2-point function of the transverse traceless part of the stress tensor in the background (\ref{bgdII}) is identical to that of empty AdS.

\begin{flushleft}
{\bf Numerical solution for $\mathbf{\Om}$ in d=3 and d=4}
\end{flushleft}

In this subsection we follow step 4 in section \ref{nure}, in order to compute ${\bf \Omega}(w;p/ \L)$ and from there, through (\ref{UV-asymptotics-OmII}), we show how to extract the renormalized response function
$\Hat\Om\sub{0}(p)$. We choose units such that $\Lambda=1$ and we numerically solve equation (\ref{masterII}) for $\Omega$ as a function of $w$ as $p$ varies. We do that for the cases $d=3$ and $d=4$. The numerical analysis for these two examples is not performed in complete detail as the purpose here is to outline the method through a simple but non-trivial example. A complete analysis\footnote{Complete in the sense of many more data points considered. In the examples of this section, only the large momentum $p$ behavior of $\Hat\Omega\sub{0}(p)$ in equation (\ref{UV-asymptotics-OmII}) is studied numerically. On the other hand, in the IHQCD example that follows, the behavior of $\Hat\Omega\sub{0}(p)$ for the whole range of $p$ is fully examined.}  is carried out in the next and more interesting example, in section \ref{hoqcd}, which concerns a geometry that belongs to the class of IHQCD.

\begin{figure}[!h]
\centering
\includegraphics[scale=0.84]{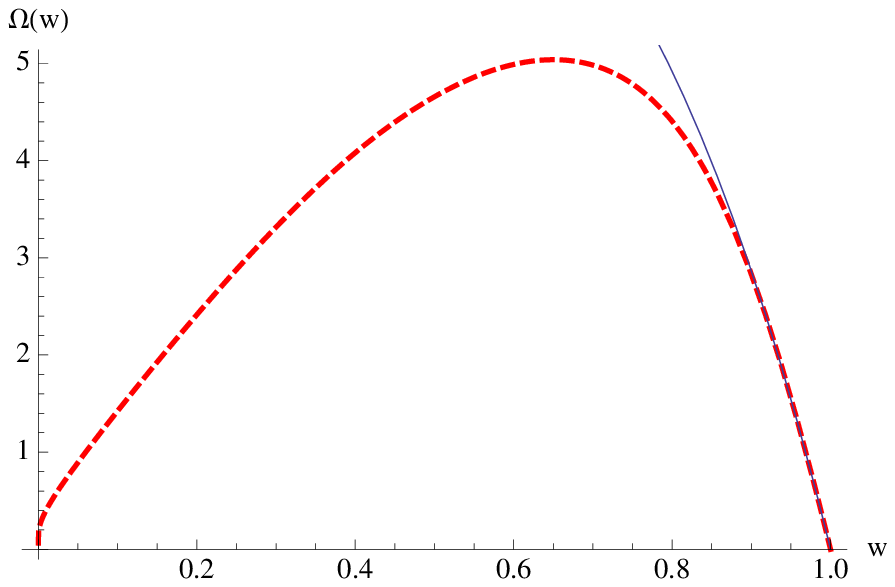}
\includegraphics[scale=0.86]{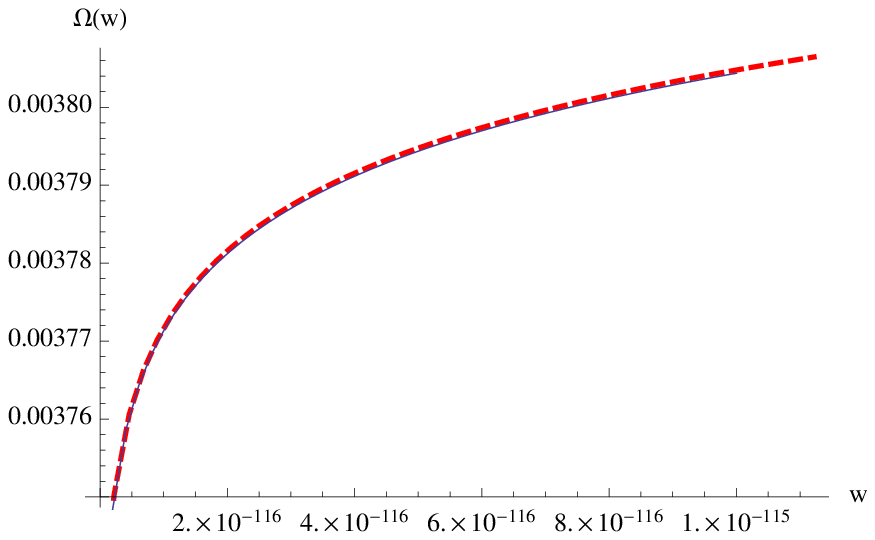}
\caption{The numerical result (red dashed line) for $\Om$ of equation (\ref{masterII}) as a function of $w$ for $d=4$ and $\frac{p}{\L}=\frac{p_{0}}{\Lambda}=10$ and $\L=1$. The geometry belongs to the class of confining geometries with a finite radial range ($w \in [0,1]$). 
{\bf Left panel}: The numerical solution is plotted for the whole range of $w$ and is superimposed in the IR (i.e. $w\to 1$) with the asymptotic solution (\ref{IR-reg-II}) (blue curve), which provides the IR boundary condition for the numerical shooting.
{\bf Right panel}: The same numerical solution is plotted in the far UV region (i.e. $w \ll 1$), and is compared with the UV asymptotic solution (\ref{UV-asymptotics-OmII}) (blue curve) with $\Hat{\Om}_{(0)}(p_0)=\log(0.707p_0)\approx 1.96$. In fact, it has been checked that for large momenta $p \gg \L$, the UV fitting is always achieved by $\Hat\Om_{(0)}(p)= \log(0.707p)$ (see (\ref{OIIplarge}) and right panel in fig. \ref{OmIId3}). Thus, as expected at large momenta, the numerical result reduces to the conformal limit (see (\ref{responseI}) for $k=\D-d/2=0$).}
\label{OmIId4}
\end{figure}

\begin{figure}[!h]
\centering
\includegraphics[scale=0.84]{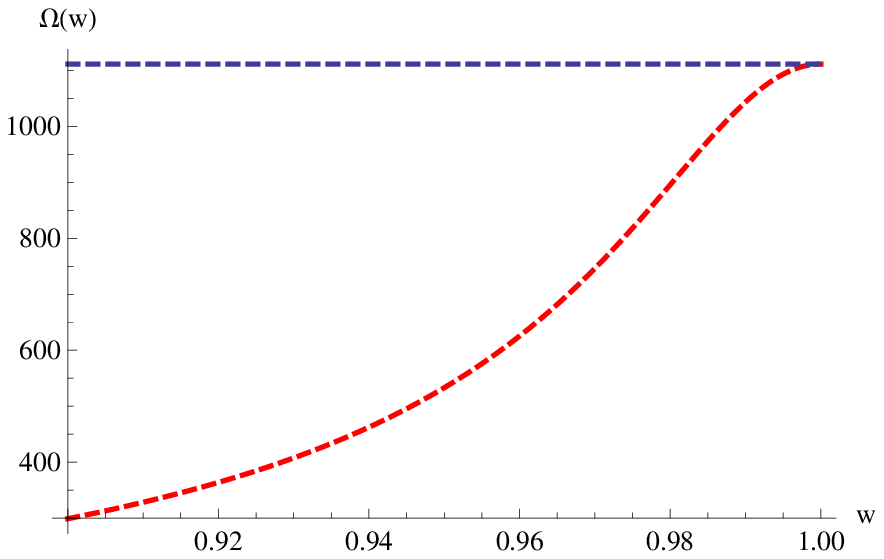}
\includegraphics[scale=0.86]{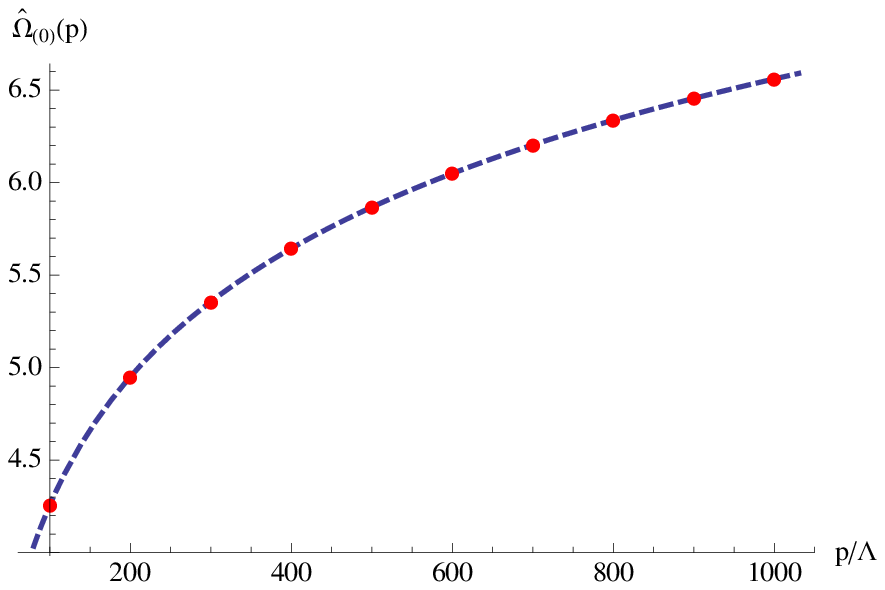}
\caption{{\bf Left panel}: The numerical solution (red dashed line) for $\Omega$ of equation (\ref{masterII}) for $d=3$ and $\frac{p}{\L}=100$ is compared with the IR asymptotic solution (blue curve), which provides the IR boundary condition for the numerical shooting. The geometry belongs to the class of confining geometries with a finite radial range ($w \in [0,1]$). The $d=3$ case is special as, according to (\ref{IR-reg-II}), in the IR it tends to the constant $\frac{p^2}{9 \L}$ in contrast to the $d=4$ case where it tends to zero (see fig. \ref{OmIId4}). {\bf Right panel}: The renormalized response function $\Hat\Om_{(0)}(p)$ extracted from the numerical solution is plotted for large momenta $p \gg \L $. The red dots
refer to $\frac{p}{\L}=100,\,200,\,..., \,1000$ and they are determined numerically by an appropriate UV fitting analogous to the right panel of fig. \ref{OmIId4}. The blue dashed curve is given by equation $\Hat\Om_{(0)}(p)=\log(0.707p)$ (see (\ref{OIIplarge})) and it evidently fits the numerically obtained $\Hat\Om_{(0)}(p)$ very accurately. Thus, as expected at large momenta, the numerical result reduces to the conformal limit (see (\ref{responseI}) for $k=\D-d/2=0$).}
\label{OmIId3}
\end{figure}

The numerical results for $\Hat\Om\sub{0}(p)$ are summarized in figures \ref{OmIId4} and \ref{OmIId3} where we show the IR and the UV behavior of $\Omega$ and extract the coefficient $\Hat\Om\sub{0}(p)$ for a few values of the momentum $p$. In particular, for large values of $p$ such that $p \gg \L$, we find
\be \label{OIIplarge}
\Hat\Om\sub{0}(p) \approx \, \log(0.707p), \,\,\, p\gg \L, \,\,\, d=3,4.
\ee
which agrees with the empty AdS result (\ref{responseI}) with $k=\Delta-d/2=0$, since $\Delta=d/2$ for this example. Therefore, the numerical results reproduce the conformal limit at large momenta as expected.

\subsection{Toy holographic QCD}
\label{hoqcd}

Our last example is a background with warp factor
\be\label{bgdIII}
A(z)=-\frac{1}{2}\L^2 z^2+\log \left(\frac{L}{z}\right),
\ee
where again $\L$ is a constant and we consider only $d=4$ in this example. As the previous example, this background is confining according to the Wilson loop test, with confinement scale $\L$ \cite{Gursoy:2007er}. From the relations (\ref{ficon}) we determine
\be\label{fip}
\phi_B'=\pm \L \sqrt{6(3+\L^2 z^2)},
\ee
together with
\be \label{WIH}
W(z)=-\frac{6}{L}e^{\frac{1}{2} \L^2 z^2} \left(1+\L^2 z^2 \right),
\ee
and
\be
V(z)=-\frac{3 e^{\L^2 z^2}}{L^2} \left(4+ 5\L^2 z^2+3 \L^4 z^4\right).
\ee
Moreover, the asymptotic form of the background scalar field takes the form 
\be
\f_B(z)=3\sqrt{2}\left(\L z+\frac{1}{18}z^3\L^3+\co(z^5)\right).
\ee
This expansion can be used to determine the asymptotic form of the superpotential
\be
W(\f_B)=-\frac{6}{L}-\frac{1}{2L}\f_B^2+\co(\f_B^4),
\ee
and of the scalar potential
\be
V(\f_B)=-\frac{12}{L ^2}-\frac{3}{2L^2}\f_B^2+\co(\f_B^4).
\ee
Comparing these with (\ref{W-asymptotics}) and (\ref{V-asymptotics}) respectively we conclude that this background describes a deformation of the dual CFT by a dimension 3 scalar operator and with deformation parameter proportional to $\L$. This is therefore exactly analogous to the GPPZ flow of $\cn=4$ super Yang-Mills \cite{Girardello:1998pd}, which was analyzed using the Riccati form of the fluctuation equations in \cite{Papadimitriou:2004rz}.

\begin{flushleft}
{\bf Riccati equations}
\end{flushleft}

The Riccati equations (\ref{master}) for this background are 
\bea\label{masterIII}
&&w\pa_w\mathbf{E}-\mathbf{E}^2-\left(4+3w^2\right)\mathbf{E}+\frac{p^2}{\L^2}w^2=0,\NO\\
&&w\pa_w\mathbf{\Om}-\mathbf{\Om}^2-\left(4+3w^2-\frac{4}{1+w^2}+\frac{6}{3+w^2}\right)\mathbf{\Om}+\frac{p^2}{\L^2}w^2=0,
\eea
where we have introduced again $w:=\L z$, and here $\mathbf{E}:= EL e^{-\L^2z^2/2}$ and $\mathbf{\Om}:= \Om L e^{-\L^2z^2/2}$.

\begin{flushleft}
{\bf IR asymptotic solutions}
\end{flushleft}

In the IR the Riccati equation (\ref{masterIII}) for $\bf{E}$ becomes
\bea
&&w\pa_w\mathbf{E}-\mathbf{E}^2-3w^2\mathbf{E}+\frac{p^2}{\L^2}w^2\approx 0,
\eea
and again $\bf{\Om}$ satisfies an identical equation.
The regular IR asymptotic solution is given by
\be\label{IRqcd}
\mathbf{E}\sim\mathbf{\Om}\sim \frac{p^2}{3\L^2}.
\ee

\begin{flushleft}
{\bf UV asymptotic solutions}
\end{flushleft}

The asymptotic UV solutions of (\ref{masterIII}) are easily found to be of the form
\bea
&&\mathbf{E}=\frac12 p^2 z^2+\frac14 p^2\left(\frac12 p^2+3\L^2\right)z^4\log(z^2\m^2)+\co(z^4),\NO\\
&&\mathbf{\Om}=-\frac{p^2L^2}{2}e^{-2A}\log (z^2\m^2)+\co(z^2),
\eea
or equivalently
\bea\label{OUVqcd}
&&EL=\frac{p^2L^2}{2}e^{-2A}+\frac{p^2L^2}{4}\left(\frac{p^2L^2}{2}e^{-2A}+\frac16\f_B^2\right)\log(z^2\m^2)+e^{-4A}L\Hat E\sub{4}(p)+\cdots,\NO\\
&&\Om L=-\frac{p^2L^2}{2}e^{-2A}\log (z^2\m^2)+e^{-2A}L\Hat\Om\sub{2}(p)+\cdots.
\eea
These expansions for $E$ and $\Om$ are identical to those for GPPZ flow derived in \cite{Papadimitriou:2004rz} (See eqs. 3.57-3.58).

\begin{flushleft}
{\bf Exact solution for $\mathbf{E}$}
\end{flushleft}

The equation for $\mathbf{E}$ in (\ref{masterIII}) can again be solved exactly. The general solution can be written 
in the form
\be\label{Edd}
\mathbf{E}=-\frac{w j'_c(w)}{j_c(w)},
\ee 
where\footnote{A nicer representation of this solution can be found in appendix \ref{generald}.} 
\be \label{jc}
j_c(w):=c\, G_{1,2}^{2,0}\left(-\frac{3 w^2}{2}\; 
\begin{array}{|c}
  1-\frac{p^2}{6\L^2} \\
  0,2 \\
\end{array}
\right)+\, w^4\, _1F_1\left(\frac{p^2}{6\L^2}+2;3;\frac{3 w^2}{2}\right),
\ee
and $G_{1,2}^{2,0}$ is the Meijer function and $_1F_1$ is Kummer's confluent hypergeometric function of the first kind. The integration constant $c$ is determined by the regularity condition (\ref{IRreg}) to be 
\be \label{c}
c=\frac{8}{9\G\left(2+\frac{p^2}{6\L^2}\right)}e^{-\frac{i\p p^2}{6\L^2}}.
\ee
This exact result, together with the expansion for $\mathbf{E}$ in (\ref{OUVqcd}) leads to
the renormalized response function  
\be \label{EIIIUV4}
\boxed{
LE\sub{4}=\frac18p^2(p^2+6\L^2)\left[\j\left(-1-\frac{p^2}{6\L^2}\right)
-\p\cot\left(\frac{\p p^2}{6\L^2}\right)\right]z^4,}
\ee
where $\psi$ is the digamma function. In appendix \ref{generald} we derive the analogous result for arbitrary even dimension $d$. As a crosscheck, we note that this expression should approach the empty AdS result given in equation (\ref{responseI}) for $d=4$ as $\Lambda \rightarrow 0$. The asymptotic expansion of the digamma function 
\be
\psi(-z)=\pi \cot(\pi z)+\log(z)+O(1/z),\,\,\, z \rightarrow \infty, 
\ee
implies that the right limit  
\be \label{checkEIII}
LE\sub{4} \sim \frac{1}{8} z^4p^4 \log(p^2), \,\,\, p \gg \L, 
\ee
is indeed recovered.

The physical content of (\ref{EIIIUV4}) can be extracted by using the identity
\be
\psi(1-y)-\psi(y)=\pi \cot(\pi y),
\ee 
and the expansion 
\be\label{idps}
\psi(2+y)=-\gamma+\sum_{n=0}^{\infty} \frac{1}{n+1}-\sum_{n=0}^{\infty} \frac{1}{n+2+y}.
\ee
we can rewrite (\ref{EIIIUV4}) as
\be\label{E-poles}\boxed{
L\Hat E\sub{4}(p) =-\frac{1}{8} \sum_{n=0}^{\infty} \frac{p^2 (p^2+6\L^2)}{n+2+\frac{p^2}{6 \L^2}},}
\ee
where we have dropped the contact terms. This means that the Lorentzian correlator, where $p^2<0$ for timelike momenta, has poles at 
\be\label{linspec}
p^2=-6\L^2  m,\quad m=2,3,\ldots. 
\ee
Such a linear spin-2 spectrum is in agreement with what was found in e.g. \cite{Gursoy:2007er} (see formulas (6.12) and (6.17) for $a=2$) and \cite{Kiritsis:2011yn} (see formulas (6.22) and (6.23)).

\begin{flushleft}
{\bf Numerical solution for $\mathbf{\Om}$}
\end{flushleft}

In this subsection we follow step 4 in section \ref{nure}, in order to extract ${\bf \Omega}(w;p/\L)$. We choose units such that $\Lambda=1$ and we introduce a convenient dimensionless function $g$ by rescaling $\bf{\Om}$ as
\be \label{g}
{\bf \Om}=\frac{p^2}{3\L^2}g(p,w).
\ee 
From (\ref{masterIII}) we see that $g(w)$ satisfies the equation   
\be \label{deg}
w\pa_w g-\frac{p^2}{3\L^2}g^2-\left(4+3w^2-\frac{4}{1+w^2}+\frac{6}{3+w^2}\right)g+3w^2=0,
\ee
and it is subject to the IR boundary condition (\ref{IRqcd}) which in the $g$ variable implies
\be \label{gbc}
g \rightarrow 1 \,\,\,\,\, \mbox{as}\,\,\,\,\,\, w \rightarrow \infty.
\ee
The utility for $g$ is that it has the same IR boundary condition for any momentum $p$ and hence checking its asymptotics becomes easier.

\begin{figure}[h!]
\centering
\includegraphics[scale=1.2]{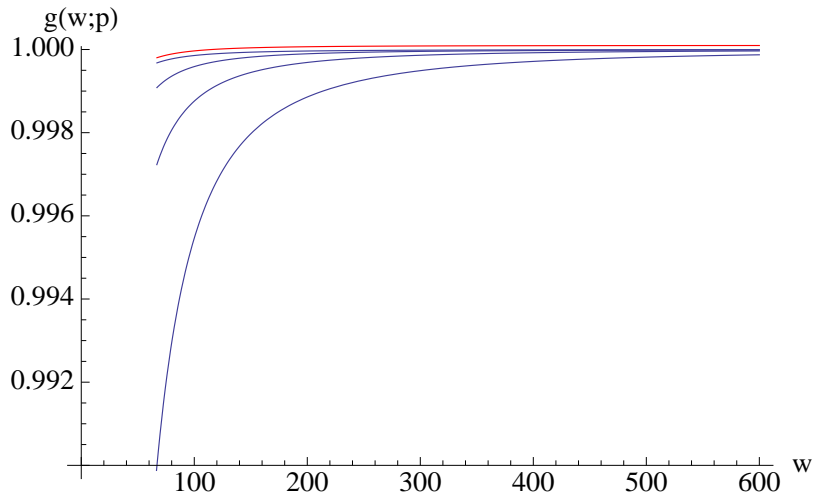}
\caption{Plot of the numerical solution for $g(w;p)$ as a function of $w=z \L$ with $\L=1$ in the IR region for several fixed momenta $p$. The IR shooting point $w_{in}$ is $w_{in}=2000$. As $p$ increases the curves move upwards and they all eventually tend to unity deep in the IR. Here we used the values $p=0.1$ (red curve), $1, \,  5, \, 10$ and  $20$.}
\label{gIRfig}
\end{figure}

\begin{figure}[h!]
\centering
\includegraphics[scale=0.85]{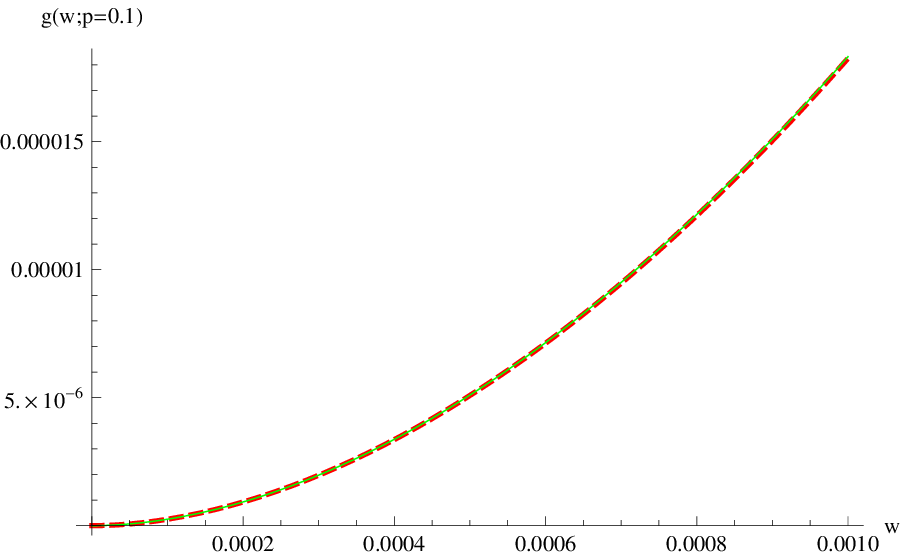}
\includegraphics[scale=0.85]{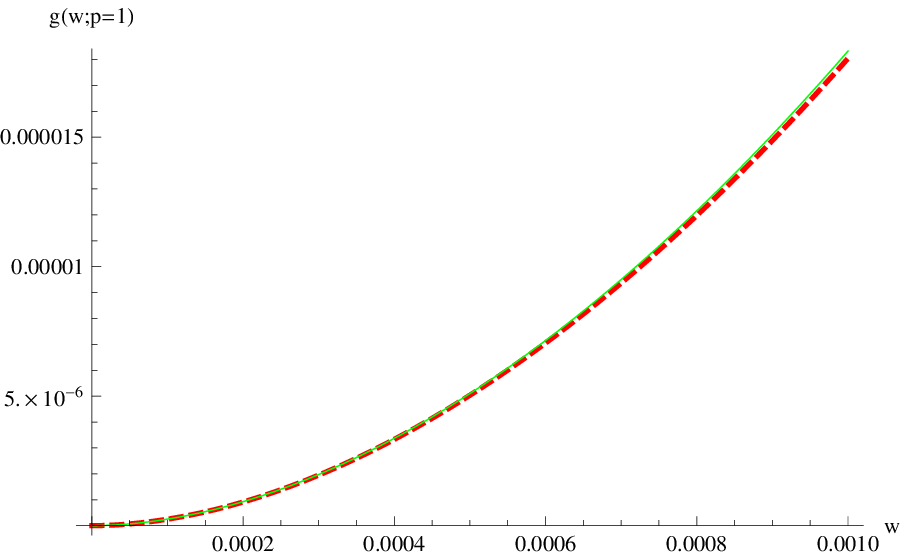}
\includegraphics[scale=0.85]{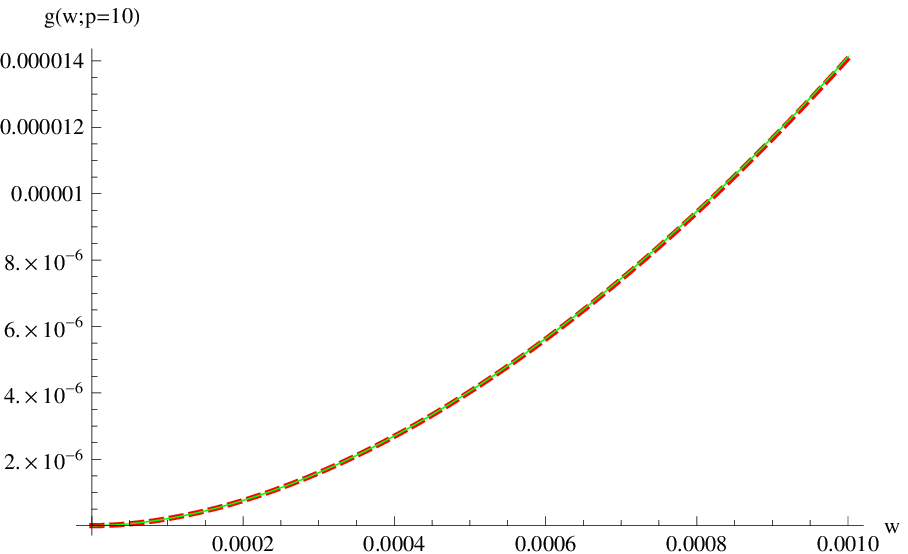}
\includegraphics[scale=0.85]{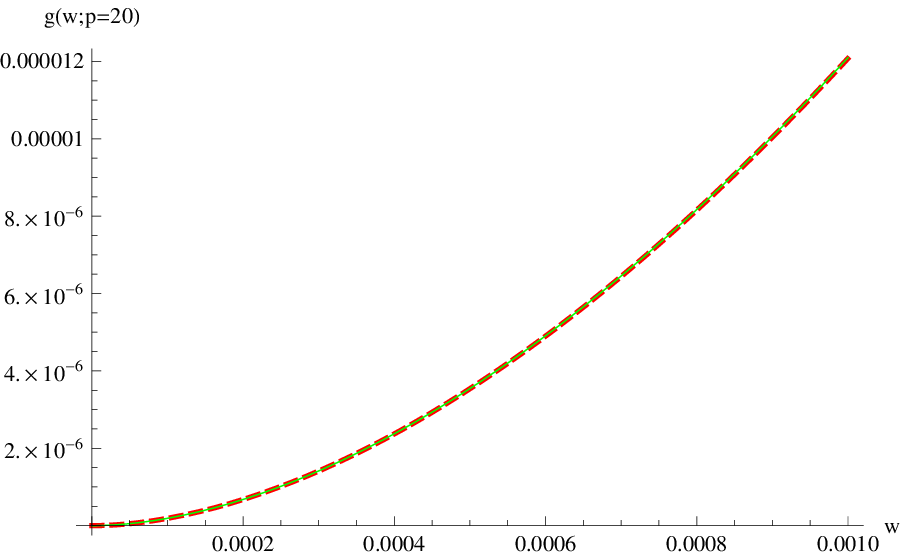}
\caption{The same numerical solution for $g(w;p)$ as in fig. \ref{gIRfig} is plotted in the UV region for several fixed momenta $p$. The red dashed curves are the numerical solutions and the green curves are the UV asymptotic solutions in (\ref{guv}). In particular, for $p=0.1$ (upper left panel), $p=1$ (upper right), $p=10$ (lower left panel), and $p=20$ (lower right) the corresponding values of $\Hat\Om_{(2)}(p)$ coefficient are  $\Hat\Om_{(2)}(p)=-0.008, \,\,-0.8,\,\,-220$ and $-1150$. Evidently, as $p$ increases $\Hat\Om_{(2)}(p)$ increases. The precise dependence of $\Hat\Om_{(2)}(p)$ on $p$ is examined in fig. \ref{c2}.}
\label{guvfig}
\end{figure}

In order to extract the QFT information, we need to specify numerically the coefficient $\Hat\Om_{(2)}$ appearing in (\ref{OUVqcd}). In terms of the $g$ variable, the UV asymptotics are
\be \label{guv}
g_{UV}\sim \frac{3\L^2}{p^2} \left( \Hat\Om_{(2)}\left(p\right) - p^2 \log(z) \right)z^2,
\ee 
where $ \Hat\Om_{(2)}\left(p\right)$ is the renormalized response function to be determined. The procedure is now straightforward. Shooting from the IR using the regularity condition (\ref{gbc}) for several $p$'s, we fit the solution at small $w$ using (\ref{guv}) and extract $\Hat\Om_{(2)}$. The numerical results for $g(w;p)$ are summarized in figures \ref{gIRfig} and \ref{guvfig} where we show the IR and UV behaviors of $g(w;p)$ and we extract the coefficient $\Hat\Om_{(2)}$ for various values of the momentum $p$.
\begin{figure}[h!]
\centering
\includegraphics[scale=1.2]{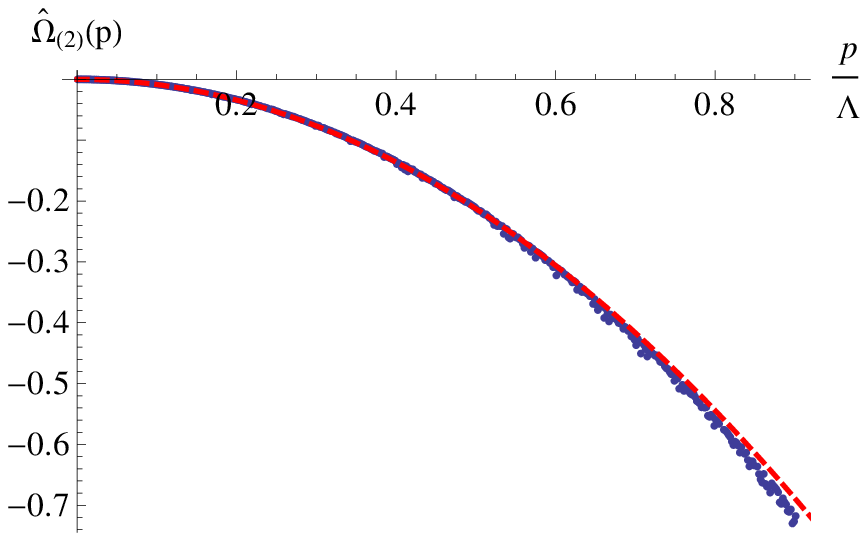}
\includegraphics[scale=1.2]{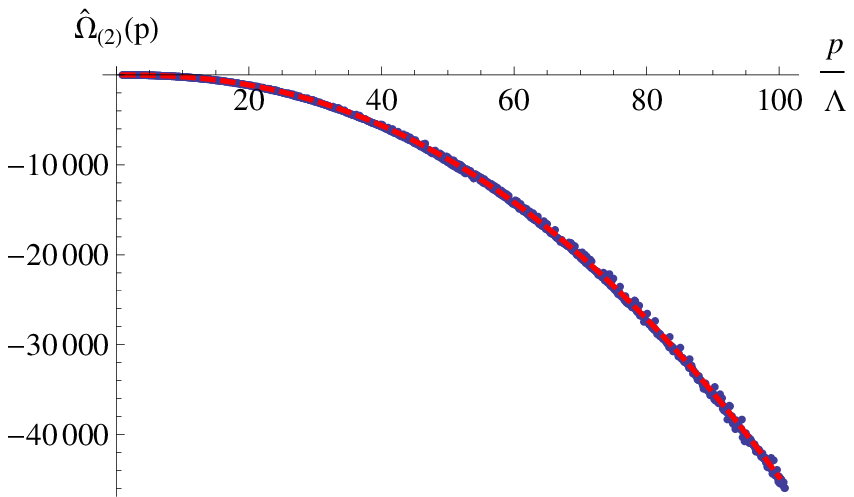}
\caption{The numerical result (blue dots) for the renormalized response function $\Hat\Om_{(2)}$ as a function of the momentum $p$. $p$ is measured in units of the confinement scale $\L$. The background geometry belongs to the class of IHQCD. The {\bf upper panel} refers to small momenta $(\frac{p}{\L} \in [0,1])$. The red dashed curve is the fitting of $\Hat\Om_{(2)}$ for small momenta. It is achieved via the function $\Hat\Om_{(2)}(p)=-0.85\, p^2$ in units of $\L=1$.
The {\bf lower panel} refers to large momenta $(\frac{p}{\L}\in [1,100])$. The red dashed curve is the fitting of $\Hat\Om_{(2)}$ for large momenta. It is achieved via the function $\Hat\Om_{(2)}(p)=-1.05\, p^2 \log( 0.707 p)$ in units of $\L=1$. This fitting matches the empty AdS behavior given by the left equation (\ref{responseI}) for $\D=3$ for IHQCD and hence for $k=\D-d/2=1$. The numerical result for small, intermediate and large momenta and the extrapolation via the fitting function (at even larger $p$) cover the whole range for all values of momenta. This fully specifies $\Hat\Om_{(2)}(p)$. }
\label{c2}
\end{figure}

Given the numerical solution for $g(w;p)$, the renormalized response function
$\Hat\Om_{(2)}(p)$ can be extracted by comparison with the asymptotic UV solution (\ref{OUVqcd}). The result is depicted in the two plots of figure \ref{c2}. The plots zoom in the small and large momentum regions, relative to the confinement scale $\L$. In the $p \gg \L$ limit, the theory is expected to be in the de-confined phase and hence the result should match the empty AdS result. Indeed, according to the second panel of the figure, the following equation
\be\label{IIIlargep}
\Hat\Om_{(2)}(p) \approx -1.05p^2\log(0.707p), \,\,\,\,\ p\gg \L=1,
\ee
fits the data very accurately as predicted by the empty AdS behavior given by the left equation in (\ref{responseI}) with $\D=3$ (for IHQCD), i.e. with $k=\Delta-d/2=1$. Away from this limit, a deviation from the conformal behavior of AdS is expected. In particular, according to the upper panel of fig. \ref{c2}, in the opposite limit $p \ll \L$ the following fitting is achieved
\be \label{IIIsmallp}
\Hat\Om_{(2)}(p) \approx -0.85p^2, \,\,\,\,\ p\ll \L=1.
\ee
The combination of the two panels of figure \ref{c2} is the main result of this section. It explicitly demonstrates the numerical procedure outlined in section \ref{nure} for the computation of the renormalized 2-point functions by using first order Riccati equations.

\section{Conclusions \& summary of results}
\label{discussion}
\setcounter{equation}{0}

In this paper we have demonstrated the utility of the Riccati form of the fluctuation equations that determine the holographic 2-point functions in the context of bottom up scalar-gravity models. 

\begin{itemize}

\item Using the Riccati form of the fluctuation equations we were able to provide a general criterion, \ref{IRreg}, for the infrared regularity of the scalar and tensor fluctuations around asymptotically AdS Poincar\'e domain wall backgrounds, and to prove in general (lemma \ref{lemma}) that provided the conditions of the holographic $c$-theorem hold, scalar and tensor fluctuations are either both singular or both regular in the infrared. These results together greatly simplify the classification of backgrounds according to their IR singularities. Four classes of IR geometries were studied in detail is section \ref{class}.

\item We provided a simplified recipe for the numerical computation of the renormalized 2-point functions. The fact that the Riccati equations are equations directly for the kernel or response functions allows us to bypass the arbitrary sources, which is a major advantage relative to the traditional method based on the second order linear fluctuation equations, especially when the 2-point functions can only be computed numerically. The recipe includes five steps which are outlined in section \ref{nure}.
 
\item In section \ref{examples} we applied this recipe to three different backgrounds, namely exact AdS and two confining geometries. One of the two confining backgrounds corresponds to a VEV of the dual scalar operator of dimension $\D=d/2$ and it is analogous to the Coulomb branch flow of $\cn=4$ super Yang-Mills. The second confining background corresponds to a deformation by a dimension 3 operator, analogous to the GPPZ flow of $\cn=4$ super Yang-Mills, and is also in the class of IHQCD backgrounds. The response function $\Hat \Om_{(2\D-d)}(p)$ for the scalar fluctuations is obtained numerically as a function of the momentum following the recipe outlined in section \ref{nure}. It is found that at large momenta, the response functions $\Hat\Om_{(2\D-d)}(p)$ behaves as in the empty AdS case with the right conformal weight $\D$ for each case. In the two confining cases, $\Hat \Om_{(2\D-d)}(p)$ is found to deviate from the AdS result at small momenta (see fig. \ref{c2}), as it should be expected. The numerically computed response function $\Hat\Om_{(2\D-d)}(p)$ determines all renormalized 2-point functions, except from the transverse traceless 2-point function of the stress tensor, through (\ref{2-point-fns-ren}). The 2-point function of the transverse traceless part of the stress tensor is determined by the response function $\Hat E_{(d)}(p)$, which we were able to compute analytically in all three examples. In the first confining background which corresponds to a VEV by a dimension $d/2$ operator we find that the non-analytic part of $\Hat E_{(d)}(p)$ is identical to the AdS result, which implies that the transverse traceless 2-point function of the stress tensor only differs from the AdS result by contact terms proportional to the scalar VEV. In the IHQCD background, however, the non-analytic part of the response function $\Hat E_{(d)}(p)$ differs from the AdS result. From the response function we extract the spin-2 spectrum, which turns out to be linear (see (\ref{linspec})). Moreover, in appendix \ref{largex} we evaluate the large distance behavior of the response function in position space, and it is found that as expected it falls exponentially at a rate proportional to the confinement scale.

\end{itemize}

In a follow up work we plan to apply our method to a wider class of models and backgrounds, to include gauge fields, an axion, and finite temperature. We believe our algorithm greatly simplifies the calculation of holographic 2-point functions in general, but especially when the latter can only be obtained numerically.  We hope that this program will be useful for applications of AdS/CFT to QCD, condensed matter and other areas of physics.

\section*{Acknowledgments}

IP is funded by the JAE-DOC program of the Consejo Superior de Investigaciones Cient\'ificas and the European Social Fund under the contract JAEDOC068. This work has also been supported by the ESF Holograv Programme, the Spanish Ministry of Economy and Competitiveness under grant FPA2012-32828, Consolider-CPAN (CSD2007-00042), the Spanish MINECO's ``Centro de Excelencia Severo Ochoa'' Programme under grant SEV-2012-0249, as well as by the grant HEPHACOS-S2009/ESP1473 from the C.A. de Madrid. AT is supported in part by the Belgian Federal Science Policy Office through the Interuniversity Attraction Pole P7/37, by FWO-Vlaanderen through projects G011410N and G020714N, by the Vrije Universiteit Brussel through the Strategic Research Program ``High-Energy Physics'' and by the VUB Research Council.  AT would like to acknowledge N. Callebaut, M. Jarvinen, R. Meyer, F. Nitti, J. Vanhoof and H. Zhang for informative and stimulating discussions.

\appendix

\renewcommand{\thesection}{\Alph{section}}
\renewcommand{\theequation}{\Alph{section}.\arabic{equation}}

\section*{Appendix}
\setcounter{section}{0}

\section{Exact transverse traceless 2-point function of the stress tensor for even d}
\label{generald}
\setcounter{equation}{0}

The generalization of (\ref{masterIII}) for arbitrary $d$ reads
\be \label{masterIIIEd}
w\pa_w\mathbf{E}-\mathbf{E}^2-\left(d+(d-1)w^2\right)\mathbf{E}+\frac{p^2}{\L^2}w^2=0,
\ee
whose general solution can be written again in the form (\ref{Edd}) with
\begin{align} 
\label{jcd}
j_{c}(w):= U\left(\frac{p^2}{2 ( d-1) \L^2}, 1-\frac d2, \frac{d-1}{2}  w^2\right) +
c_d \cdot {}_1F_1\left(\frac{p^2}{2 ( d-1) \L^2},1-\frac d2 , \frac{d-1}{2}  w^2\right),
\end{align}
where $_1F_1$ and $U$ are Kummer's confluent hypergeometric functions of the first and second kind respectively, and $c_d$ is an arbitrary integration constant. Since $_1F_1$ grows exponentially for large argument while $U$ goes to zero, the regularity condition (\ref{IRreg}) requires that $c_d=0$. We therefore obtain
\be\label{}\boxed{
\mathbf{E}=-\frac{w\pa_w U\left(\frac{p^2}{2 ( d-1) \L^2}, 1-\frac d2, \frac{d-1}{2}  w^2\right)}{U\left(\frac{p^2}{2 ( d-1) \L^2}, 1-\frac d2, \frac{d-1}{2}  w^2\right)}.}
\ee
This solution is completely general and valid for both even and odd $d$. However, the asymptotic expansion of this expression for even and odd $d$ are different, and hence so is the corresponding renormalized response function $\Hat E\sub{d}$.  For even $d$ we can use the expansion 
\bea
&&U\left(\frac{p^2}{2 ( d-1) \L^2}, 1-\frac d2, \frac{d-1}{2}  w^2\right)=
\frac{\G\left(\frac d2\right)}{\G\left(\frac d2+\frac{p^2}{2 ( d-1) \L^2}\right)}\left(1-\frac{p^2}{2 ( d-2) \L^2}w^2+\cdots\right)\\
&&+\frac{(-1)^{1+\frac d2}\left(\frac{d-1}{2}\right)^{\frac d2} w^d}{\G\left(1+\frac d2\right)\G\left(\frac{p^2}{2 ( d-1) \L^2}\right)}
\left(\log w^2+\psi\left(\frac d2+\frac{p^2}{2 ( d-1) \L^2}\right)-\psi\left(1+\frac d2\right)-\psi(1)+\cdots\right),\NO
\eea
from which we deduce that the non-analytic part of the response function $\mathbf{E}$ is
\be\label{Ed}\boxed{
L\Hat E\sub{d} = \frac{2(-1)^{\frac d2}\left(\frac{d-1}{2}\right)^{\frac d2}\G\left(\frac d2+\frac{p^2}{2 ( d-1) \L^2}\right)}{\G\left(\frac d2\right)^2\G\left(\frac{p^2}{2 ( d-1) \L^2}\right)}\psi\left(\frac d2+\frac{p^2}{2 ( d-1) \L^2}\right).}
\ee
For $d=4$ this agrees with (\ref{EIIIUV4}). By expanding the digamma function as in the $d=4$ case we see the spin-2 spectrum remains linear for any even $d$.

\section{Long distance behavior of the stress tensor 2-point function for IHQCD}
\label{largex}
\setcounter{equation}{0}

In this section we extract the long distance behavior of the stress tensor 2-point function in the example of section \ref{hoqcd}. Starting with (\ref{E-poles}) and Fourier transforming to position space we have
\be
\int d^4p  e^{i {p\cdot x} }  L\Hat E\sub{4}(p) =-\frac{1}{8} \sum_{n=0}^{\infty} \int d^4{ p} e^{i p\cdot x} \frac{p^2 (p^2+6\L^2)}{n+2+\frac{p^2}{6 \L^2}},
\ee
Performing the angular integrations using 
\be
\int e^{i {p\cdot x}} d\Om_3=4\p \int_0^{\pi} d\th e^{i p x \cos\th} \sin^2\th=4 \pi^2 \frac{J_1(p x)}{px}, 
\ee
we obtain
\begin{align}
&\int d^4p  e^{i p\cdot x } L\Hat E\sub{4}(p) =-\frac{\p^2}{2} \square^2 (6\L^2-\square) \frac1x\sum_{n=0}^{\infty} \int_0^{\infty}  \frac{dp J_1(p x)}{n+2+\frac{p^2}{6 \L^2}}  \notag\\
&=-\frac{\p^2}{2}6\L^2 \square^2 (6\L^2-\square)  \sum_{n=0}^{\infty} \left( \frac{1}{6(2+n)\L^2 x^2}-\frac{K_1\left(\sqrt{6(2+n)}\L x\right)}{\sqrt{6(2+n)}\L x} \right).
\end{align}
Dropping the first term in the parentheses since it only gives a contact term and noting that 
\be
\square \left(\frac{K_1\left(\sqrt{6(2+n)}\L x\right)}{\sqrt{6(2+n)}\L x}\right)=6(2+n)\L^2 \left(\frac{K_1\left(\sqrt{6(2+n)}\L x\right)}{\sqrt{6(2+n)}\L x}\right),
\ee
where 
\be
\square = \pa_x^2 +\frac3x\pa_x,
\ee
is the radial part of the Laplacian in $d=4$, we arrive at the following expression for the position space 2-point function
\be\label{Ex}
\boxed{
\int d^4p  e^{i p\cdot x } L\Hat E\sub{4}(p) = -\frac{\p^2}{2} (6\L^2)^4 \sum_{n=0}^{\infty} (n+1)(n+2)^2
\frac{K_1\left(\sqrt{6(2+n)}\L x\right)}{\sqrt{6(2+n)}\L x}.}
\ee
Formula (\ref{Ex}) expresses the response function in position space as an infinite sum of Bessel functions. In particular, in the large separation limit this sum is dominated by the $n=0$ term, and hence for $\L x>>1$
\be\label{ExL}
\boxed{
\int d^4p  e^{i p\cdot x } L\Hat E\sub{4}(p) \sim -2\p^2(6\L^2)^4\sqrt{\frac{\p}{2}}
\frac{e^{-2\sqrt{3}\L x}}{(2\sqrt{3}\L x)^{3/2}}.}
\ee
As expected, the 2-point function decays exponentially at large distances with rate proportional to the confinement scale $\L$. This result can be easily generalized to arbitrary dimension using the response function computed in appendix \ref{generald}.

\bibliographystyle{jhepcap}
\bibliography{dw2ptfns}

\end{document}